\documentclass[pra,showpacs,showkeys,amsfonts,amsmath]{revtex4}
\usepackage{bm}
\usepackage{graphicx}
\RequirePackage{mathptm}
\newtheorem{thm}{Theorem}

\newtheorem{lem}{Lemma}

\newtheorem{rem}{Remark}
\numberwithin{equation}{section}

\newcommand{\ip}[2]{\langle \,{#1},\,{#2}\,\rangle}
\newcommand{\W}[4]{\begin{cases}
#1 ,&#2\\[2mm]
#3 ,&#4
\end{cases}}
\newcommand{\ro}{\varrho}

\newcommand{\la}{\lambda}
\newcommand{\las}[1]{\lambda_{#1}}

\newcommand{\al}{\alpha}

\newcommand{\ee}{\varepsilon}

\newcommand{\vf}{\varphi}

\newcommand{\I}{\openone}
\newcommand{\conj}[1]{\overline{#1}}

\newcommand{\ket}[1]{|{#1}\rangle}
\newcommand{\bra}[1]{\langle {#1} |}

\newcommand{\Z}{\mathbb Z}
\newcommand{\C}{\mathbb C}
\newcommand{\R}{\mathbb R}
\newcommand{\M}{\mathbb M}
\newcommand{\B}{\mathbb B}
\newcommand{\D}{\mathbb D}
\newcommand{\PP}{\mathbb P}

\newcommand{\fE}{\mathcal E}
\newcommand{\cA}{{\mathcal A}}
\newcommand{\cB}{{\mathcal B}}
\newcommand{\cM}{\mathcal M}
\newcommand{\cP}{\mathcal P}
\newcommand{\cJ}{\mathcal J}
\newcommand{\cK}{\mathcal K}
\newcommand{\cW}{\mathcal W}
\newcommand{\K}{\mathbb K}
\newcommand{\cT}{\mathcal T}

\newcommand{\tr}{\mathrm{tr}\,}
\newcommand{\ptr}[1]{\mathrm{tr}_{#1}}

\newcommand{\mr}[1]{\mathrm{#1}}

\begin{document}
\title{Measurement - induced qudit geometric discord}
\author{Piotr
{\L}ugiewicz,  Andrzej Frydryszak \footnote{ andrzej.frydryszak@ift.uni.wroc.pl} and Lech Jak{\'o}bczyk
\footnote{ ljak@ift.uni.wroc.pl} }
 \affiliation{Institute of Theoretical Physics\\ University of
Wroc{\l}aw\\
Plac Maxa Borna 9, 50-204 Wroc{\l}aw, Poland}
\begin{abstract}
We study the measurement-induced geometric discord based on the trace norm and generalize some properties  known for qutrits to  qudits. Previous preliminary  results for bipartite qutrit systems (i.e. $d=3$ systems) are here  strictly proved for arbitrary $d$. Present study supports observations, coming also from other approaches, that  systems with $d\geq 3$ show similar behaviour when quantum correlations are concerned, but there is pronounced difference between $d=2$ and $d=3$. Qubit systems are exceptionally simple. Underlying geometry of state spaces and related Lie groups are responsible for that.
\end{abstract}
 \maketitle
\section{Introduction}
Quantum correlations in finite dimensional quantum systems became an area of research to which  a lot
of effort is directed. In this context, the first nontrivial and basic system for understanding quantum correlations is the two - qubit system.
It was  firstly studied with respect to  quantum entanglement which in turn is
the most studied quantum correlation by now. Results obtained for qubit system  can be relatively
easily generalized to some extend to a qubit-qudit case, where qudit is a $d$-level system.
Two qubit system is also the first one used to study  more general quantum correlations. Last decade
brought more deep understanding of the notion of 'quantumness' and more sophisticated differentiation
of types of correlations in states of compound quantum systems. While pure states can be uncorrelated
or entangled, mixed states exhibit more subtle hierarchy of non-classical correlations. Today's
classification distinguishes following types of states with non-classical correlations:  nonlocal, steerable, entangled, generally quantumly correlated \cite{ABC}. For pure states these attributes are synonymous.

In recent years we witness a lot of activity related to study of the
quantum discord as a measure of quantum correlations (see e.g.
\cite{OZ,MBC,AFA,CBR,AFCA,MPM,BCFA,HJM}. This notion is very
general, catching difference of the quantum and classical character
of correlations in compound systems, but difficult to calculate even
for the two-qubit system. To have more efficient tool there were
introduced various modified measures of quantum correlations like:
geometric measures, measurement-induced geometric measures,
measurement induced informational measures, entanglement activation
measures, unitary response measures, coherence based measures and
recoverability measures \cite{ABC}. Within the  geometric measures
there appears subclassification related to the distance used in
definition. The one firstly studied was the Hilbert-Schmidt
distance, convenient in calculations, but not contractive under
completely positive trace preserving maps \cite{Piani, Ozawa}, and
therefore not suitable to define a bona fide measure of quantum
correlations. For that reason geometric measures based on trace norm
(Schatten 1-norm) are more proper, but alas, less easy to handle.

However, as various  works including the present one show,  properties of bipartite systems with
$d\geq 3$ change  strongly. There, one meets new situation also characterized by recently found
obstruction that  there is no finite set of criteria  of separability for two-qutrit states \cite{SKO}.
Therefore, as one gets outside the Peres-Horodecki's necessary and
sufficient PPT-criterion for qubit-qubit and qubit-qutrit systems\cite{P, HH1} the analysis becomes
harder and one has to rely on other tools.

In the previous work we  studied  qutrit systems
\cite{JFL} and we obtained some preliminary results for limited class of
states, which allowed generalization to the case of quantum correlations in bipartite qudit system.
Here we look for  the widest set of states allowing the similar procedure of computation  of the measurement-induced quantum geometric discord as found in previous work, the  analysis is based on  general assumptions and strict proofs. What is important, the geometric discord we use, is based on the trace norm, which is much more difficult to compute, then the one based on the Hilbert-Schmidt distance. The measurement-induced quantum geometric discord is defined as the minimal disturbance induced by any projective measurement on the subsystem of compound quantum system, computed
using the trace distance. Such measure can be compared with the
standard geometric discord equal to the distance from a given state
to the set of classical-quantum states \cite{DVB}. Generally, the
measurement-induced geometric discord dominates the geometric
discord and these two quantities are equal if a distance used in
definition is the Hilbert-Schmidt distance \cite{Luo}. For the trace
distance, the measurement-induced geometric discord and geometric
discord coincide only for qubits \cite{Na, RSI}. The
measurement-induced geometric discord based on trace norm  is a bona
fide correlation measure  and allows to obtain explicit results for
various families of states. As it is known, even for bipartite qubit
system the minimization can be solved analytically only for the
Hilbert-Schmidt distance \cite{DVB} and this basically follows from
the geometry of the qubit state space. In the case of trace distance
it is possible for the limited set of families of mixed states
\cite{Cic}. For qutrits and higher dimensional qudits situation is
even more hard.
\par
The main goal of the present work is to provide strict analysis of
selected properties of the measurement-induced quantum geometric
discord for arbitrary $d$. The paper is organized as follows. In the
next section we provide some background information and fix the
notation to describe generic qudit system. Then some geometrical
properties of the $\R^{d^2-1}$ related to the structure of the
algebra $\mr{su(d)}$ are recalled, as well as definition of the
one-sided measurement-induced quantum geometric discord is
commented. In Section III we describe locally maximally mixed states
and study the form of the disturbance of such states induced by
local measurements. We find general relevant lower bound for the
trace of a square of such disturbance and show for which family of
states it is saturated. To obtain exact formula for the trace-norm
quantum discord further simplifications are necessary. The Section
IV contains analysis of the  two-qutrit system which serves as a
guiding example for finding what simplifications should be assumed
to obtain exact result for the trace norm measurement - induced
geometric discord for qudits. Such generalization is given in the
Section V, where we show that such discord can be obtained without
performing minimization procedure for relevant equivalence classes of
correlation matrices. This is crucial, due to the fact that the minimization procedure for $d>2$
is not known and for general case, presumably, not computable at all.
In the Section VI there are discussed
instructive examples of two-qudit states illustrating subtle points
of the previously proven theorems. We conclude with some comments on
complexity and effectiveness of calculation procedure for
measurement-induced quantum discord. Some technical points as well
as detailed discussion of the quantum correlations in the family of
two-qutrit states with diagonal orthogonal correlation matrices  are
shifted to the Appendices.

\section{Preliminary notions}
The basic notions relevant for a description of qudit systems where
already introduced in \cite{JFL}. For the reader convenience we
recall them in this section, including the discussion of adjoint
representations of the group $\mr{SU(d)}$ and the corresponding
geometry of the parameter space of $d$ - level quantum systems. We
stress also fundamental difference between qubits ($d=2$) and higher
dimensional qudits.
\subsection{Qudits}
Let us start with description of $d$ - level ($d\geq 3$) quantum
systems (qudits). The corresponding Hilbert space equals to $\C^{d}$
and observables are given by hermitian elements of full matrix
algebra $\M_{d}(\C)$. It is convenient to use as a basis in
$\M_{d}(\C)$ the hermitian generators of $\mr{su(d)}$ algebra and
the identity matrix $\I_{d}$. Let $\las{1},\ldots,\las{d^{2}-1}$  be
the generators of $\mr{su(d)}$ algebra.   The matrices $\las{j}$
satisfy
\begin{equation*}
\tr\, \las{j}=0,\quad \tr\, (\las{j}\las{k})=2\,\delta_{jk},\;
j,k=1,\ldots,d^{2}-1
\end{equation*}
and
\begin{equation}
\las{j}\las{k}=\frac{2}{d}\,\delta_{jk}\,\I_{d}
+\sum\limits_{l}\,(\hat{d}_{jkl}+i\,\hat{f}_{jkl})\,\las{l}\label{lajlak}
\end{equation}
where the structure constants $\hat{d}_{jkl}$ and $\hat{f}_{jkl}$
are given by
\begin{equation}
\hat{d}_{jkl}=\frac{1}{4}\,\tr\,([\las{j},\las{k}]_{+}\,\las{l})\label{d}
\end{equation}
and
\begin{equation}
\hat{f}_{jkl}=\frac{1}{4i}\,\tr\,([\las{j},\las{k}]\,\las{l}).\label{f}
\end{equation}
Using the structure constants (\ref{d}) and (\ref{f}) one can
introduce the following "star" and "wedge" products in a real linear
space  $\R^{d^{2}-1}$. For $n,\, m\in \R^{d^{2}-1}$ we define
\begin{equation}
(n\star
m)_{j}=\sqrt{\frac{d(d-1)}{2}}\,\frac{1}{d-2}\,\sum\limits_{k,l}\,\hat{d}_{jkl}n_{k}m_{l}
\end{equation}
and
\begin{equation}
(n\wedge
m)_{j}=\sqrt{\frac{d(d-1)}{2}}\,\frac{1}{d-2}\,\sum\limits_{k,l}\,\hat{f}_{jkl}n_{k}m_{l}
\end{equation}
Let us note, that above two formulas do not cover the $d=2$ case. In particular, $\hat{d}_{ijk}$
constants vanish and the $\star$-product is trivial for qubits.
Let $\la=(\las{1},\ldots,\las{d^{2}-1})$ and
\begin{equation}
\ip{n}{\la}=\sum\limits_{j}n_{j}\las{j}
\end{equation}
then taking into account (\ref{lajlak}), we obtain
\begin{equation}
\ip{n}{\la}\ip{m}{\la}=\frac{2}{d}\,\ip{n}{m}\I_{d}+\frac{1}{d^{\prime}}\,\ip{n\star
m}{\la}+\frac{i}{d^{\prime}}\,\ip{n\wedge m}{\la}, \label{product}
\end{equation}
where
\begin{equation*}
d^{\prime}=\sqrt{\frac{d(d-1)}{2}}\,\frac{1}{d-2}.
\end{equation*}
The set of observables i.e. the subspace of hermitian elements of $\M_{d}(\C)$, forms a Jordan algebra with respect to the Jordan product
\begin{equation*}
A\circ B=\frac{1}{2}(AB+BA)
\end{equation*}
which for matrices
\begin{equation*}
A=a_{0}\,\I_{d}+\ip{a}{\la},\quad B=b_{0}\I_{d}+\ip{b}{\la}
\end{equation*}
where $a,\, b\in \R^{d^{2}-1}$, is given by
\begin{equation}
A\circ B=\left(a_{0}b_{0}+\frac{2}{d}\ip{a}{b}\right)\I_{d}+\frac{1}{d^{\prime}}\ip{b_{0}a+a_{0}b+a\star b}{\la}.
\label{jordan}
\end{equation}
The set $\fE_{d}$ of all states of $d$ - level system can be parametrized as
follows (see e.g. \cite{BK})
\begin{equation}
\ro=\frac{1}{d}\,\left(\I_{d}+d^{\prime\prime}\,\ip{n}{\la}\right),\quad
n\in \R^{d^{2}-1},\label{state}
\end{equation}
where
\begin{equation*}
d^{\prime\prime}=\sqrt{\frac{d(d-1)}{2}}
\end{equation*}
and the components of the vector $n$ are
\begin{equation*}
n_{j}=\frac{d}{\sqrt{2d(d-1)}}\,\tr\,(\ro\,\la_{j}),\quad
j=1,\ldots,d^{2}-1.
\end{equation*}
The matrix (\ref{state}) is hermitian and has a unit trace. To describe a quantum state, the matrix $\ro$ have to be
positive-definite
and this condition is not easy to characterize in terms of the vector $n$. However the pure states given
by one-dimensional projectors
can be fully described. Using (\ref{product}), one can check that $\ro$ given by (\ref{state}) satisfies
$\ro^{2}=\ro$ if and only if
\begin{equation*}
\ip{n}{n}=1\quad\text{and}\quad n\star n=n.
\end{equation*}
As it is well known, the case of qubits ($d=2$) is very special.
Since in that case in the formula  (\ref{product})
the star product is absent, the set of observables forms the
Jordan algebra which is called spin factor (see e.g. \cite{HOS}).
Moreover, the set of states can be easily characterized in terms of
the vectors $n$: it is the unit ball in $\R^{3}$ and the pure states
correspond to the unit sphere. For qutrits and higher order qudits the $n\star n=n$ condition becomes
nontrivial and prevents the simple geometrical characterization of one-qudit state space. At the moment
some more specific results, but not simple, are known for qutrits only \cite{GSSS}.
\par
Consider now two qudits $\cA$ and $\cB$. It is convenient to
parametrize the set of states of composite system as follows
\begin{equation}
\ro=\frac{1}{d^{2}}\left(\I_{d}\otimes
\I_{d}+d^{\prime\prime}\,\ip{x}{\la}\otimes \I_{d}+\I_{d}\otimes
d^{\prime\prime}\,\ip{y}{\la}
+\sum\limits_{k=1}^{d^{2}-1}\ip{\cK\,e_{k}}{\la}\otimes\ip{e_{k}}{\las{k}}\right)
\label{9state}
\end{equation}
where $x,\, y\in \R^{d^{2}-1}$ and $\{e_{k}\}_{k=1}^{d^{2}-1}$ are the vectors of canonical orthonormal basis of $\R^{d^{2}-1}$. Notice that
\begin{equation*}
x_{j}=\frac{d}{\sqrt{2d(d-1)}}\,\tr\,(\ro\,\las{j}\otimes\I_{d}),\quad
y_{j}=\frac{d}{\sqrt{2d(d-1)}}\,\tr\,(\ro\,\I_{d}\otimes \las{j})
\end{equation*}
and the correlation matrix $\cK$ has elements
\begin{equation*}
\cK_{jk}=\frac{d^{2}}{4}\,\tr\,(\ro\las{j}\otimes\las{k}).
\end{equation*}
The parametrization (\ref{9state}) is chosen is such a way, that the marginals
$\ptr{\cA}\ro$ and $\ptr{\cB}\ro$ are given by the vectors $x$ and $y$ as in (\ref{state}).
\subsection{Adjoint representation of $\mr{SU(d)}$ and geometry of  $\R^{d^{2}-1}$}
Let us now discuss briefly the adjoint representation of the group $\mr{SU(d)}$. Let $U\in \mr{SU(d)}$ and define
the matrix $R(U)$ by
\begin{equation*}
\ip{R(U)m}{\la}\equiv U\ip{m}{\la}U^{\ast},\quad m\in \R^{d^{2}-1},
\end{equation*}
or
\begin{equation*}
U\,\la_{j}\,U^{\ast}=\sum\limits_{k}R(U)_{kj}\,\la_{k}.
\end{equation*}
The matrix elements of $R(U)$ are given by
\begin{equation*}
R(U)_{jk}=\frac{1}{2}\,\tr\,(U\,\la_{k}\,U^{\ast}\,\la_{j})
\end{equation*}
By the mapping $R$, to each element $U\in \mr{SU(d)}$ there correspond real orthogonal matrix $R(U)\in \mr{SO(d^{2}-1)}$ and let
\begin{equation*}
\mr{G}(d)\equiv R(SU(d))\subset \mr{SO(d^{2}-1)}
\end{equation*}
Since the dimension of the group $\mr{SU(d)}$ is $d^{2}-1$, and the
dimension of $\mr{SO(d^{2}-1)}$ equals to
$\frac{1}{2}\,(d^{2}-1)(d^{2}-2)$, the matrices $R(U)\in \mr{G}(d)$
form only a very small part of the group $\mr{SO(d^{2}-1)}$. In
particular, $\mr{G}(3)$ contains linear transformations which leave
invariant  inner product  in $\R^{8}$ and cubic invariant
$\ip{n\star n}{n}$ \cite{Mac}.
Again the case when $d=2$ is exceptional. The group $\mr{G}(2)$
exactly equals to $\mr{SO(3)}$ for which the $SU(2)$ is the double covering group. Therefore using
adjoint group, for $d=2$ we have full control over $\R^{d^{2}-1}=\R^3$ space.
For $d=3$ we have discrepancy in dimensions: $8$ for $G(3)$ and $28$ for $SO(8)$. This only wideness
for higher $d$. Such geometrical effect makes analysis of two - qubit system much simpler then in
higher dimensions.
\par
Consider now the covariance properties of the  star and wedge products defined on the linear space $\R^{d^{2}-1}$. It follows from the fact that
 $\hat{f}_{jkl}$ and $\hat{d}_{jkl}$ are invariant tensors that
\begin{equation}
 Vm\,\wedge\, Vn=V(m\, \wedge\, n),\quad Vm\,\star \,Vn=V\,(m\star n),
\end{equation}
for all $V\in \mr{G}(d)$. Define also the matrices
\begin{equation}
(\Delta_{j})_{kl}=\hat{d}_{jkl},\quad (F_{j})_{kl}=\hat{f}_{jkl}.
\end{equation}
Now for all $V\in \mr{G}(d)$
\begin{equation}
V^{T}\Delta_{j}V=\sum\limits_{k=1}^{d^{2}-1}V_{jk}\,\Delta_{k}
\end{equation}
and similarly
\begin{equation}
V^{T}F_{j}V=\sum\limits_{k=1}^{d^{2}-1}V_{jk}\,F_{k}.
\end{equation}
 For the further applications we will need the following
property of the star product.
\begin{lem}
The equality
\begin{equation}
\sum\limits_{k=1}^{d^{2}-1}A\,e_{k}\,\star\,B\,e_{k}=0,\quad\text{(zero vector)}\label{sumastar}
\end{equation}
is satisfied if and only if
\begin{equation*}
\tr\, (A^{T}\Delta_{j}\,B)=0,\quad\text{for all}\quad j=1,\ldots,d^{2}-1
\end{equation*}
\end{lem}
Proof:\\
 Notice that
\begin{equation*}
A\,e_{k}\,\star\, B\,e_{k}=d^{\prime}\,\sum\limits_{j}(A^{T}\,\Delta_{j}\,B)_{kk}\,e_{j}
\end{equation*}
so
\begin{equation*}
\sum\limits_{k}A\,e_{k}\,\star\,B\,e_{k}=d^{\prime}\,\sum\limits_{k,j}(A^{T}\,\Delta_{j}\,B)_{kk}\,e_{j}=
d^{\prime}\,\sum\limits_{j}\,\tr\,(A^{T}\,\Delta_{j}\,B)\, e_{j}
\end{equation*}
$\Box$
\par
Since $\tr\, \Delta_{j}=0,\, j=1,\ldots, d^{2}-1$, we have in particular
\begin{equation}
\sum\limits_{k=1}^{d^{2}-1} e_{k}\,\star e_{k}=0\label{stare}
\end{equation}
and equality (\ref{stare}) is true for any orthonormal basis of $\R^{d^{2}-1}$.
\subsection{ Measurement - induced qudit geometric discord}
When a bipartite system $\cA\cB$ is prepared in a state $\ro$ and we
perform local measurement on the subsystem $\cA$, almost all states
$\ro$ will be disturbed due to such measurement. The \textit{one-sided (measurement - induced)
geometric discord} is defined as the minimal disturbance induced by
any projective measurement $\PP_{\cA}$ on subsystem $\cA$
\cite{DVB}. In the standard approach Hilbert - Schmidt norm is used to measure a distance
in the set of states and the corresponding quantum discord is denoted by $D^{M}_{2}$. Here we prefer
to choose a distance given by the trace norm and define quantum discord $D^{M}_{1}$ as  \cite{Paula}
\begin{equation}
D^{M}_{1}(\ro)=\frac{d}{2(d-1)}\,\min\limits_{\PP_{\cA}}\,||\ro-\PP_{\cA}(\ro)||_{1},\label{trdisc}
\end{equation}
where
\begin{equation*}
||A||_{1}=\tr\,|A|.
\end{equation*}
On the other hand, $D^{M}_{2}$ is defined as
\begin{equation}
D^{M}_{2}(\ro)=\frac{d}{d-1}\,\min\limits_{\PP_{\cA}}\,||\ro-\PP_{\cA}(\ro)||_{2}^{2}\label{HSdisc}
\end{equation}
where
\begin{equation*}
||A||_{2}=\sqrt{\tr\, A^{\ast}A}
\end{equation*}
In the case of qudits, local projective measurement $\PP_{\cA}$ is
given by the one-dimensional  projectors $P_{1},\, P_{2},\ldots,\,
P_{d}$ on $\C^{d}$, such that
\begin{equation*}
P_{1}+P_{2}+\cdots + P_{d}=\I_{d},\quad
P_{j}P_{k}=\delta_{jk}\,P_{k}
\end{equation*}
and $\PP_{\cA}=\PP\otimes \mr{id}$, where
\begin{equation}
 \PP(A)=P_{1}\,A\, P_{1}+P_{2}\,A\, P_{2}+\cdots +P_{d}\,A\,P_{d}.\label{PP}
\end{equation}
One - dimensional projectors $P_{k}$ can be always chosen as
\begin{equation*}
P_{k}=U\,P_{k}^{0}\,U^{\ast}\quad\text{for some}\quad U\in \mr{SU(d)},
\end{equation*}
where $P_{k}^{0}=\ket{\vf_{k}}\bra{\vf_{k}}$ and $\{\vf_{k}\}$ is a
standard orthonormal basis in $\C^{d}$. If $\PP_{0}$ is the mapping (\ref{PP}) given by $P_{k}^{0}$, then
\begin{equation}
\PP_{0}(A)=\mr{diag}\, (a_{11},\,a_{22},\, \ldots,\, a_{dd})
\end{equation}
Define a real orthogonal
projector $\cP$ on $\R^{d^{2}-1}$
\begin{equation}
\ip{\cP m}{\la}=\PP(\ip{m}{\la}),\quad m\in \R^{d^{1}-1},\label{realproj}
\end{equation}
or
\begin{equation*}
\cP_{jk}=\frac{1}{2}\,\tr\, (\PP(\las{j})\las{k}).
\end{equation*}
If   $\cP_{0}$ denotes such projector corresponding to $\PP_{0}$,
then
\begin{equation*}
\cP=V\,\cP_{0}\,V^{T},\quad V\in \mr{G}(d)
\end{equation*}
Notice that the matrices $\las{j}$ for $j=k^{2}-1,\; k=2,3,\ldots,d$ are diagonal, whereas remaining $\las{j}$ have zero diagonal elements, so
\begin{equation*}
\PP_{0}(\las{j})=\W{\las{j}}{j=k^{2}-1}{0}{j\neq k^{2}-1}
\end{equation*}
Thus $\cP_{0}$ projects on $d-1$ dimensional subspace and only non zero matrix elements of $\cP_{0}$ are
\begin{equation}
(\cP_{0})_{k^{2}-1,k^{2}-1}=1,\quad k=2,3,\ldots,d.
\end{equation}
Define also orthogonal complements to $\cP_{0}$ and $\cP$
\begin{equation}
\cM_{0}=\I-\cP_{0},\quad \cM=\I-\cP. \label{M}
\end{equation}
Obviously $\cM=V\,\cM_{0}\,V^{T},\; V\in \mr{G(d)}$ and
\begin{equation*}
\mr{dim}\,\mr{Ran}\,\cM_{0}=\mr{dim}\,\mr{Ran}\,\cM=d(d-1).
\end{equation*}
Notice that only in the case of qubits, where
$\mr{G(3)}=\mr{SO(3)}$, the projectors $\cM$ run over the set of all
orthogonal projectors with a fixed dimension. When $d\geq 3$, this
set is a proper subset of all such projectors and it causes the
minimization problem below difficult to solve.
\par
Let us compute now the disturbance of the state (\ref{9state})
induced by measurement $\PP_{\cA}$. Since $\PP_{\cA}$ acts only on
subsystem $\cA$, we obtain
\begin{equation}
\ro-\PP_{\cA}(\ro)=\frac{1}{d^{2}}\,\big[
d^{\prime\prime}\ip{\cM x}{\la}\otimes
\I_{d}+\sum\limits_{k}\ip{\cM\cK\,e_{k}}{\la}\otimes
\ip{e_{k}}{\la}\big] \label{dif}
\end{equation}
Let $S(\cM)$ denotes the right hand side of equation (\ref{dif}). Then
\begin{equation}
D^{M}_{1}(\ro)=\frac{d}{2(d-1)}\,\min\limits_{\cM}\,\tr\,\sqrt{Q(\cM)}
\end{equation}
where $Q(\cM)=S(\cM)S(\cM)^{\ast}$ and the minimum is taken over all matrices $\cM$ corresponding to a measurements on subsystem $\cA$.
Similarly
\begin{equation}
D^{M}_{2}(\ro)=\frac{d}{d-1}\,\min\limits_{\cM}\,\tr\, Q(\cM)
\end{equation}
\section{Locally maximally mixed states}
Let us consider the class of locally maximally mixed states i.e.
such states $\ro$ that
\begin{equation}
\ptr{\cA}\ro=\frac{\I_{d}}{d},\quad \ptr{\cB}\ro=\frac{\I_{d}}{d}.\label{MMM}
\end{equation}
In the parametrization (\ref{9state}) this property corresponds to $x=y=0$ and we have
\begin{equation}
\ro=\frac{1}{d^{2}}\left(\I_{d}\otimes \I_{d}+\sum\limits_{j=1}^{d^{2}-1}\ip{\cK e_{j}}{\la}\otimes \ip{e_{j}}{\la}\right)
\label{lmm}
\end{equation}
Let $\K^{(d)}$ be the set of correlation matrices corresponding to
(\ref{lmm}). The set $\K^{(d)}$ is convex, contains zero matrix and
$\K^{(d)}\subset \B_{2}$, where
\begin{equation}
\B_{2}=\left\{A\in \M_{d^{2}-1}(\R)\;:\; ||A||_{2}\leq \frac{d}{2}\sqrt{d^{2}-1}\right\}
\end{equation}
This last property follows from the condition $\tr\, \ro^{2}\leq 1$, since for the states (\ref{lmm})
\begin{equation*}
\tr \, \ro^{2}=\frac{1}{d^{2}}+\frac{4}{d^{4}}\,||\cK||_{2}^{2}.
\end{equation*}
One can check that pure states in this class, which are in fact
maximally entangled, are defined by correlation matrices lying on
the boundary of the ball $\B_{2}$, but not every such matrix
corresponds to some state, so  detailed characterization of the set
$\K^{(d)}$ is a real problem and the general solution is not known.
\par
Let $\cK\in \K^{(d)}$, then
\begin{equation}
S(\cM)=\frac{1}{d^{2}}\,\sum\limits_{j=1}^{d^{2}-1}\,\ip{\cM\cK\,e_{j}}{\la}\otimes
\ip{e_{j}}{\la},
\end{equation}
and
\begin{equation}
\begin{split}
Q(\cM)=\frac{1}{d^{4}}\,\bigg[&\,\frac{4}{d^{2}}\,\sum\limits_{j}\,\ip{\cM\cK\,e_{j}}{\cM\cK\,e_{j}}\,
\I_{d}\otimes \I_{d}
+\frac{2}{d\,d^{\prime}}\,\sum\limits_{j}\,\ip{\cM\cK\,e_{j}\star \cM\cK\,e_{j}}{\la}\otimes\I_{d}\\
&+\frac{2}{d\,d^{\prime}}\,\sum\limits_{j,k}\,\ip{\cM\cK\,e_{j}}{\cM\cK\,e_{k}}\I_{d}\otimes \ip{e_{j}\star e_{k}}{\la}
+\frac{1}{d^{\prime 2}}\,\sum\limits_{j,k}\,\ip{\cM\cK\,e_{j}\star \cM\cK\,e_{k}}{\la}\otimes \ip{e_{j}\star e_{k}}{\la}\\
&-\frac{1}{d^{\prime
2}}\sum\limits_{j,k}\,\ip{\cM\cK\,e_{j}\wedge \cM\cK\,e_{k}}{\la}\otimes
\ip{e_{j}\wedge e_{k}}{\la}\,\bigg]
\end{split}\label{QM}
\end{equation}
If $\cK$ is general correlation matrix it is a difficult task to
compute the spectrum of $Q(\cM)$ and obtain an analytic expression
for the measure of discord. But we are  able to
 find the universal lower bound for $D_{1}^{M}$ and $D_{2}^{M}$.  Observe that
\begin{equation*}
\tr\,
Q(\cM)=\frac{4}{d^{4}}\,\sum\limits_{j=1}^{d^{2}-1}\ip{\cM\cK\,e_{j}}{\cM\cK\,e_{j}}=\frac{4}{d^{4}}\,\tr\,
(\cK\cK^{T}\cM),
\end{equation*}
so
\begin{equation*}
D_{2}^{M}(\ro)=\frac{d}{d-1}\,\min\limits_{\cM}\;\tr\,Q(\cM)=\frac{4}{d^{3}(d-1)}\,\min\limits_{\cM}\;
\tr\,(\cK\cK^{T}\cM).
\end{equation*}
Since
\begin{equation*}
\tr\,\sqrt{Q(\cM)}\geq \sqrt{\tr\,Q(\cM)},
\end{equation*}
we also have
\begin{equation*}
D_{1}^{M}(\ro)\geq\frac{1}{d(d-1)}\sqrt{\min\limits_{\cM}\;\tr\,(\cK\cK^{T}\cM)}.
\end{equation*}
Thus we need  a lower bound for the quantity given by
$\min\limits_{\cM}\;\tr\,(\cK\cK^{T}\cM)$ and this bound can be
obtained by applying the following general result:
\begin{lem}
Let $A$ be a non - negative operator acting on the space
$\R^{n_{0}}$, with eigenvalues $\mu_{1}\geq \mu_{2}\geq \cdots \geq
\mu_{n_{0}}$. Let $P$ be any orthogonal projector on $\R^{n_{0}}$,
such that  $\tr\, P=m_{0}$ and $m_{0}< n_{0}$. Then
\begin{equation}
\min\limits_{P}\,\tr\, PA\ = \sum\limits_{j=n_{0}-m_{0}+1}^{n_{0}}\,\mu_{j}
\end{equation}
\end{lem}
Proof:\\
Let
\begin{equation*}
A=\sum\limits_{j=1}^{n_{0}}\mu_{j}E_{j}
\end{equation*}
Put
\begin{equation*}
\omega_{j}=\tr\,PE_{j},\quad j=1,\ldots,n_{0}
\end{equation*}
Then $\omega_{j}\in [0,1]$ and
$\omega_{1}+\cdots+\omega_{n_{0}}=m_{0}$. Consider the function
\begin{equation*}
f(\vec{\omega})=\sum\limits_{j=1}^{n_{0}}\mu_{j}\omega_{j},\quad
\vec{\omega}=(\omega_{1},\ldots,\omega_{n_{0}})
\end{equation*}
defined on the set
\begin{equation*}
\Omega=\{\vec{\omega}\,:\, \omega_{j}\in [0,1],\,
\sum\limits_{j=1}^{n_{0}}\omega_{j}=m_{0}\}
\end{equation*}
We are looking for the minimal value of the function
$f(\vec{\omega})$. Let $J_{0}$ and $J_{1}$ be disjoint subsets of
the set $\{1,2,\ldots,n_{0}\}$. Define the subset of $\Omega$
\begin{equation*}
\Omega_{J_{0},J_{1}}=\{\vec{\omega}\,:\,
\omega_{j}=0\quad\text{for}\quad j\in J_{0}\quad\text{and}\quad
\omega_{j}=1\quad\text{for}\quad j\in J_{1}\}
\end{equation*}
Using the method of Lagrange multipliers we obtain on
$\Omega_{J_{0},J_{1}}$
\begin{equation}
\mu_{j}-\nu=0,\quad\text{for}\quad j\in J_{0}^{c}\cap
J_{1}^{c}\quad\text{and}\quad \sum\limits_{j\in J_{0}^{c}\cap
J_{1}^{c}}\omega_{j}=m_{0}-|J_{1}|\label{La}
\end{equation}
The equations (\ref{La}) have a solution if the eigenvalues of $A$
have a proper degeneracy. In this case one can compute that
\begin{equation*}
 f(\vec{\omega})\geq
\sum\limits_{j=n_{0}-m_{0}+1}^{n_{0}}\mu_{j}\quad \text{for all}\quad \vec{\omega}\in \Omega_{J_{0},J_{1}}
\end{equation*}
If there is no proper degeneracy, we can add one point to $J_{0}$ or $J_{1}$ i.e. we pass to the sets $(J_{0}^{\prime},J_{1})$
or $(J_{0},J_{1}^{\prime})$ an repeat the above reasoning. After the finite number of steps we arrive at such pair $(J_{0},J_{1})$ that
$|J_{0}|=n_{0}-m_{0}$ and $|J_{1}|=m_{0}$. The function $f(\vec{\omega})$ assumes the smallest value if $\vec{\omega}\in \Omega_{J_{0},J_{1}}$ where
$J_{0}=\{1,\ldots, m_{0}\}$ and $J_{1}=\{n_{0}-m_{0}+1,\ldots,n_{0}\}$ and the smallest value is equal to $\sum\limits_{j=n_{0}-m_{0}+1}^{n_{0}}\mu_{j}$. $\Box$
 \vskip 4mm\noindent
 To  apply this result, take $A=\cK\cK^{T},\;
n_{0}=d^{2}-1$, and $P= \cM$, so $m_{0}=d(d-1)$. Let
 $\{\eta_{j}^{\downarrow}\}$ be the eigenvalues of $\cK\cK^{T}$
  in non- increasing  order and define
\begin{equation*}
\Xi(\cK)=\sum\limits_{j=d}^{d^{2}-1}\eta_{j}^{\downarrow}.
\end{equation*}
Since $\cM=V\cM_{0}V^{T},\; V\in \mr{G}(d)\subset \mr{SO}(d^{2}-1)$,
the minimum is taken over a proper subset of the set of all
projections on $\R^{d^{2}-1}$, so by the Lemma 2
\begin{equation*}
\min\limits_{\cM}\; \tr\,(\cK\cK^{T}\cM)\geq \Xi(\cK)
\end{equation*}
and we have:
\begin{thm}
Let $\cK\in\K^{(d)}$ and $\ro$ be the corresponding locally
maximally mixed state, then
\begin{equation*}
D_{2}^{M}(\ro)\geq
\frac{4}{d^{3}(d-1)}\;\Xi(\cK)\quad\text{and}\quad
D_{1}^{M}(\ro)\geq \frac{1}{d(d-1)}\;\sqrt{\Xi(\cK)}
\end{equation*}
In particular, when $\mr{rank}\,\cK\geq d$, the corresponding
state has non - zero quantum discord.
\end{thm}
Notice that the Theorem 1 gives an alternative justification of the
lower bound on Hilbert - Schmidt quantum discord established in
\cite{RP, HL}.
\par
This result gives only the lower bound of quantum discord. When we
consider a special case of the matrix $\cK$, we can obtain more
detailed information. Take $\cK=t\, V_{0}$, where $V_{0}\in
\mr{O(d^{2}-1)}$ and $t$ is a real parameter, such that $t\,V_{0}\in
\K^{(d)}$. Since $t\,V_{0}$ should be in $\B_{2}$
\begin{equation*}
||t\,V_{0}||_{2}^{2}=t^{2}\,(d^{2}-1)\leq \frac{d^{2}}{4}\,(d^{2}-1)
\end{equation*}
so $t$ belongs to the interval $|t|\leq d/2$, but the actual value
of $t$ depend on the choice of the matrix $V_{0}$. Notice that for
such correlation matrices
\begin{equation*}
\Xi(\cK)=t^{2}\,d(d-1)
\end{equation*}
so for the states with such $\cK$
\begin{equation*}
D_{2}^{M}(\ro)\geq\frac{4t^{2}}{d^{2}}
\end{equation*}
On the other hand
\begin{equation*}
\sum\limits_{j}\ip{\cM\cK\,e_{j}}{\cM\cK\,e_{j}}=t^{2}\,\sum\limits_{j}
\ip{\cM V_{0}\,e_{j}}{V_{0}\,e_{j}}=t^{2} \tr\,\cM=t^{2}\,d(d-1).
\end{equation*}
so $\tr\, Q(\cM)$ does not depend on local measurement matrix and to
compute quantum discord $D_{2}^{M}$ we do not need to minimize over
all $\cM$. Thus we obtain
\begin{thm}
For locally maximally mixed two - qudit states $\ro$ with the
correlation matrix $\cK=t\,V_{0},\, V_{0}\in \mr{O(d^{2}-1)}$, we
have
\begin{equation*}
D^{M}_{2}(\ro)=\frac{4t^{2}}{d^{2}}
\end{equation*}
so the lower bound for $D_{2}^{M}$ is tight.
\end{thm}
\begin{rem}
Concerning trace - norm quantum discord, we have only the lower
bound
\begin{equation*}
D_{1}^{M}(\ro)\geq \frac{|t|}{\sqrt{d(d-1)}}
\end{equation*}
The above  result shows that every state defined by the correlation
matrix $\cK=t\,V_{0}$ with $t\neq 0$, has non - zero quantum
discord. On the other hand, every such state is separable at least
for the parameters $t$ in the interval
\begin{equation*}
|t|\leq \frac{d}{4(d^{2}-1)}
\end{equation*}
as it follows from the sufficient condition of separability: the
states $\ro$ satisfying  $\tr\,\ro^{2}\leq 1/(d^{2}-1)$ are
separable \cite{GB}.
\end{rem}
\par
Now we  show that in the case $\cK=t\,V_{0}$,  the formula
(\ref{QM}) for $Q(\cM)$ can be simplified. We start with the proof
that the second term in (\ref{QM}) vanishes.
\begin{lem}
The operator $\cM$  defined by (\ref{M}) satisfies the condition
\begin{equation}
\sum\limits_{j=1}^{d^{2}-1}\cM Ve_{j}\,\star\, \cM Ve_{j}=0\label{starM}
\end{equation}
for any $V\in \mr{O(d^{2}-1)}$.
\end{lem}
Proof:\\
By Lemma 1, the condition (\ref{starM}) is equivalent to
\begin{equation}
\tr\,( (\cM V)^{T}\,\Delta_{i}\cM V)=\tr\, \cM\,\Delta_{i}=0 \quad\text{for all}\quad i=1,\ldots, d^{2}-1\label{Mdelta}
\end{equation}
First we show that
\begin{equation}
\tr\, \cM_{0}\,\Delta_{j}=0,\quad j=1,\ldots,d^{2}-1.\label{M0delta}
\end{equation}
It is enough to check that the matrix $\cP_{0}$ satisfies the
condition (\ref{M0delta}). Since $\cP_{0}$ is a projector
\begin{equation*}
\tr\,(\cP_{0}\Delta_{i})=\tr\,(\cP_{0}\Delta_{i}\cP_{0})=\sum\limits_{l=2}^{d}(\Delta_{i})_{l^{2}-1,\,l^{2}-1}=
\sum\limits_{l=2}^{d}d_{i,\,l^{2}-1,\,l^{2}-1}.
\end{equation*}
On the other hand
\begin{equation*}
d_{i,\,l^{2}-1,\,l^{2}-1}=\frac{1}{2}\,\tr\,(\las{i}\,\las{l^{2}-1}^{2}),
\end{equation*}
so
\begin{equation*}
\tr\,(\cP_{0}\Delta_{i})=\frac{1}{2}\tr(\las{i}\,\sum\limits_{l=2}^{d}\las{l^{2}-1}^{2})=
\frac{d-1}{d}\,\tr\,(\las{i}\,\I_{d})=0
\end{equation*}
since
\begin{equation*}
\las{3}^{2}+\las{8}^{2}+\cdots
+\las{d^{2}-1}^{2}=\frac{2(d-1)}{d}\,\I_{d}
\end{equation*}
Now $M=R(U)\,M_{0}\,R(U)^{T}$ for some  $U\in \mr{SU(d)}$ and
\begin{equation*}
R(U)^{T}\Delta_{i}R(U)=R(U)_{ij}\,\Delta_{j}
\end{equation*}
so we have
\begin{equation*}
\tr \,\cM\,\Delta_{i}=\tr\,(R(U)\,\cM_{0}\,R(U)^{T}\,\Delta_{i})=\tr\,(\cM_{0}\,R(U)^{T}\,\Delta_{i}\,R(U))=\sum\limits_{j}
R(U)_{ij}\,\tr\,\cM_{0}\,\Delta_{j}
\end{equation*}
and condition (\ref{Mdelta}) follows. $\Box$
\par
Next we prove that the remaining terms in the formula for $Q(\cM)$ can
be transformed such that we obtain the following result:
\begin{thm}
Let $\cK=t\,V_{0},\, V_{0}\in \mr{O(d^{2}-1)}$.
Then
\begin{equation}
Q(\cM)=\frac{t^{2}}{d^{4}}\bigg[\frac{4(d-1)}{d}\,\I_{d}\otimes
\I_{d}+\frac{2}{d}\,\I_{d}\otimes \sum\limits_{k}X_{k}\,\las{k}
+\sum\limits_{j,k}Y_{jk}\,\las{j}\otimes \las{k}\bigg]
\label{QMI}
\end{equation}
where
\begin{equation}
X_{k}=\tr\,(\cM\,V_{0}\Delta_{k}V_{0}^{T}),\quad
Y_{jk}=\tr\,(V_{0}^{T}\cM\Delta_{j}\cM V_{0}\Delta_{k}+V_{0}^{T}\cM F_{j}\cM V_{0}F_{k})
\end{equation}
\end{thm}
Proof:\\
First we consider the third term in (\ref{QM})
\begin{equation}
\sum\limits_{j,k}\ip{\cM\cK\,e_{j}}{\cM\cK\,e_{k}}\I_{d}\otimes
\ip{e_{j}\star
e_{k}}{\la}
=t^{2}\,\sum\limits_{j,k}\ip{V_{0}\,e_{j}}{\cM V_{0}\,e_{k}}\,\I_{d}\otimes
\ip{e_{j}\star e_{k}}{\la}
=t^{2}\,\sum\limits_{j,k}\ip{e_{j}}{V_{0}^{T}\cM V_{0}\,e_{k}}\,\I_{d}\otimes
\ip{e_{j}\star e_{k}}{\la}.
\label{3te}
\end{equation}
Since
\begin{equation*}
\sum\limits_{j}\ip{e_{j}}{V_{0}^{T}\cM V_{0}\,e_{k}}\,\I_{d}\otimes\ip{e_{j}\star
e_{k}}{\la}
=\I_{d}\otimes \ip{\sum\limits_{j}\ip{e_{j}}{V_{0}^{T}\cM V_{0}\,e_{k}}e_{j}\star
e_{k}}{\la}
=\I_{d}\otimes\,\ip{V_{0}^{T}\cM V_{0}\,e_{k}\star e_{k}}{\la}
\end{equation*}
the sum (\ref{3te}) is equal to
\begin{equation*}
t^{2}\,\I_{d}\otimes \sum\limits_{k}\ip{V_{0}^{T}\cM V_{0}\,e_{k}\star e_{k}}{\la}
=d^{\prime}\,t^{2}\,\I_{d}\otimes \sum\limits_{k}X_{k}\,\las{k}
\end{equation*}
To simplify  fourth and fifth terms in the formula (\ref{QM}) notice
that
\begin{equation}
\ip{\cM\cK\,e_{j}\star
\cM\cK\,e_{k}}{\la}=t^{2}\,\ip{\cM V_{0}\,e_{j}\star\,\cM V_{0}\,e_{k}}{\la}
=d^{\prime}t^{2}\,\sum\limits_{l}[(\cM V_{0})^{T}\Delta_{l}(\cM V_{0})]_{jk}\,\las{l}
\label{4te}
\end{equation}
and
\begin{equation}
\ip{\cM\cK\,e_{j}\wedge \cM\cK\,e_{k}}{\la}=t^{2}\,\ip{\cM V_{0}\,e_{j}\wedge
\cM V_{0}\,e_{k}}{\la}
=d^{\prime}t^{2}\,\sum\limits_{l}[(\cM V_{0})^{T}F_{l}(\cM V_{0})]_{jk}\las{l}
\label{5te}
\end{equation}
On the other hand
\begin{equation}
\ip{e_{j}\star
e_{k}}{\la}=d^{\prime}\sum\limits_{p}(\Delta_{p})_{jk}\las{p}\label{4s}
\end{equation}
and
\begin{equation}
\ip{e_{j}\wedge
e_{k}}{\la}=d^{\prime}\,\sum\limits_{p}(F_{p})_{jk}\,\las{p}\label{5w}
\end{equation}
Using (\ref{4te}) and (\ref{4s}) we see that the fourth term of
(\ref{QM}) is equal to
\begin{equation*}
t^{2}\sum\limits_{l,p,j,k}[(\cM V_{0})^{T}\Delta_{l}(\cM V_{0})]_{jk}(\Delta_{p})_{jk}\,\las{l}\otimes
\las{p}
=t^{2}\,
\sum\limits_{l,p}\,\tr\,(V_{0}^{T}\cM\Delta_{l}\cM V_{0}\Delta_{p})\,\las{l}\otimes
\las{p}.
\end{equation*}
Similarly, by (\ref{5te}) and (\ref{5w}) we obtain the fifth term
\begin{equation*}
-\,t^{2}\sum\limits_{l,p,j,k}[(\cM V_{0})^{T}F_{l}\cM V_{0}]_{jk}(F_{p})_{jk}\,\las{l}\otimes
\las{p}
=t^{2}\,\sum\limits_{l,p}\tr\,(V_{0}^{T}\cM F_{l}\cM V_{0}F_{p})\,\las{l}\otimes
\las{p},
\end{equation*}
where the change of sign follows from the antisymmetricity of
matrices $F_{p}$. Combining all above results we arrive at the
formula (\ref{QMI}). $\Box$
\par
 The formula (\ref{QMI}) is a starting point for further
 simplifications in order to obtain exact expression for trace -
 norm quantum discord. To find necessary conditions on the
 correlation matrices, first we will analyse  the case
 of two qutrits.
\section{The formula for $Q(\cM)$. The case of qutrits.}
Even in the case of correlation matrix $\cK=t\,V_{0}$, we can find only a lower
bound on a trace norm geometric discord. To obtain exact value of
$D^{M}_{1}$, we still need some simplifications in the formula (\ref{QMI}).  We consider first the case of two
qutrits and focus on  diagonal orthogonal matrices i.e. such matrices $I$ that $I^{2}=\I_{8}$.
The case of diagonal matrices was already considered in our previous
work \cite{JFL}, where by a direct computation we have found that
the  matrix
\begin{equation}
I_{0}=\mr{diag}\,(1,-1,1,1,-1,1,-1,1)
\end{equation}
corresponding for example to qutrit Bell state, satisfies
\begin{equation}
\sum\limits_{k}I_{0}\cM I_{0}\,e_{k}\star\,e_{k}=0.\label{I0MI0star}
\end{equation}
Under this condition, the formula for $Q(\cM)$ simplifies
considerably and one can check that
\begin{equation}
\tr\,Q(\cM)^{k}=\tr\,Q(\cM_{0})^{k},\quad k=1,\ldots,9.
\end{equation}
So it follows that the eigenvalues of $Q(\cM)$ and $Q(\cM_{0})$ are
the same (see e.g. \cite{L}) and for the states with the correlation
matrix $\cK=t\,I_{0}$ we can compute $D^{M}_{1}$ by finding the
trace norm of $\sqrt{Q(\cM_{0})}$ and we need not to minimize over
all local measurements. Unfortunately, due to the computational
complexity, this method can be applied only to limited class of
qutrit states and does not give any hints how to treat  higher
dimensional qudits.
\par
In the present analysis we reverse the reasoning and we first  look
for the condition on the arbitrary diagonal orthogonal matrix $I$
under which equality analogous to (\ref{I0MI0star}) is satisfied. It
turns out that we are able to fully characterize such matrices $I$
and to find the compact formula for $Q(\cM)$. Moreover, the analysis
can be naturally extended to arbitrary qudits.
To formulate the result, let us introduce the mapping
$\tau_{I}\,:\,\M_{3}(\C)\to\M_{3}(\C)$
\begin{equation}
\tau_{I}\left(a_{0}\I_{3}+\ip{a}{\la}\right)=a_{0}\I_{3}+\ip{Ia}{\la}
\end{equation}
Now we have:
\begin{thm}
The condition
\begin{equation}
\sum\limits_{k}I\cM I\,e_{k}\star\,e_{k}=0\label{IMIstar}
\end{equation}
is satisfied for all local measurement matrices $\cM$ if and only if
the mapping $\tau_{I}$ corresponding to the matrix $I$ is the Jordan
automorphism of the algebra $\M_{3}(\C)$.
\end{thm}
Proof:\\
Assume that $\tau_{I}$ is a Jordan automorphism. It means that
$\tau_{I}(A\circ B)=\tau_{I}(A)\circ \tau_{I}(B)$. By (\ref{jordan})
we have
\begin{equation*}
\left(a_{0}b_{0}+\frac{2}{3}\ip{a}{b}\right)\,\I_{3}+\frac{1}{\sqrt{3}}\,\ip{b_{0}Ia+a_{0}Ib+I(a\star
b)}{\la}=
\left(a_{0}b_{0}+\frac{2}{3}\ip{Ia}{Ib}\right)\,\I_{3}+\frac{1}{\sqrt{3}}\,\ip{b_{0}Ia+a_{0}Ib+Ia\star
Ib}{\la},
\end{equation*}
so the matrix $I$ satisfies $\ip{Ia}{Ib}=\ip{a}{b}$ and $I(a\star
b)=Ia\star Ib$. Since
\begin{equation*}
\tr\,((\cM I)^{T}\Delta_{k}I)=\tr\,(I^{2}\cM\Delta_{k})=\tr\,(\cM\Delta_{k})=0,\quad
k=1,\ldots, 8,
\end{equation*}
it follows that
\begin{equation*}
\sum\limits_{k}\cM Ie_{k}\star Ie_{k}=0
\end{equation*}
and
\begin{equation*}
\sum\limits_{k}I\cM Ie_{k}\star e_{k}=\sum\limits_{k}I(\cM I)e_{k}\star
I(Ie_{k})=I\,\sum\limits_{k}\cM Ie_{k}\star Ie_{k}=0
\end{equation*}
The proof that from the condition (\ref{IMIstar}) follows that
$\tau_{I}$ is a Jordan automorphism is much more involved and is
based on two lemmas below (for the proofs see Appendix).
\begin{lem}
If the condition (\ref{IMIstar}) is satisfied then for all $U\in \mr{SU(3)}$
\begin{equation}
(\tau_{I}(U\las{3}U^{\ast}))^{2}+(\tau_{I}(U\las{8}U^{\ast}))^{2}=\frac{4}{3}\,\I_{3}.\label{la3la8}
\end{equation}
\end{lem}
\begin{lem}
The equation (\ref{la3la8}) is fulfilled if and only if the mapping
$\tau_{I}\,:\, \M_{3}(\C)\to \M_{3}(\C)$ is positive.
\end{lem}
From the Lemma 4 and Lemma 5  we obtain in particular that if condition
(\ref{IMIstar}) is satisfied by some matrix $I$,  the corresponding
mapping $\tau_{I}$ is positive. Now the property that this mapping
is a Jordan automorphism follows, since $\tau_{I}$ is positive only
if the matrix $I$ has the form (see proof of Lemma 5)
\begin{equation}
I=\mr{diag}\,(\ee_{1},\,\ee_{2},\,1,\,\ee_{1}\ee_{2}\ee_{5},\,\ee_{5},\,\ee_{2}\ee_{5},\,\ee_{1}\ee_{5},\,1)\label{Ipositive}
\end{equation}
where $\ee_{1},\, \ee_{2},\, \ee_{5}\in \{1,-1\}$ and one can check
that such $I$ defines Jordan automorphism. $\Box$
\par
From the above proof it follows that there are only $8$ matrices $I$
such that the mapping $\tau_{I}$ is a Jordan automorphism. Notice
that this set can be divided into two classes:
\begin{equation}
[\I_{8}]=\{\I_{8},\, V_{1},\, V_{2},\,V_{3}\}
\end{equation}
where $V_{k}=R(W_{k}), k=1,2,3$, and $W_{k}$ are given by
\begin{equation}
W_{1}=\mr{diag}\,(1,-1,-1),\quad W_{2}=\mr{diag}\,(-1,1,-1),\quad W_{3}=\mr{diag}\,(-1,-1,1)\label{W123}
\end{equation}
and
\begin{equation}
[I_{0}]=\{I_{0},\,I_{0}V_{1},\,I_{0}V_{2},\,I_{0}V_{3}\}
\end{equation}
where
\begin{equation}
I_{0}=\mr{diag}\,(1,-1,1,1,-1,1,-1,1)\label{I0}
\end{equation}
If $I\in [\I_{8}]$, the mapping $\tau_{I}$ is an automorphism of
$\M_{3}(\C)$: identity mapping for $I=\I_{8}$ and
\begin{equation}
\tau_{I}(A)=W_{k}AW_{k},\quad I=V_{k}
\end{equation}
On the other hand, the elements of the class $[I_{0}]$ define anti -
automorphisms of $\M_{3}(\C)$. If $I=I_{0}$ then
\begin{equation}
\tau_{I}(A)=A^{T}
\end{equation}
and for $I=I_{0}V_{k}$ we have
\begin{equation}
\tau_{I}(A)=W_{k}A^{T}W_{k}
\end{equation}
\begin{thm}
If the matrix $I$ defines an automorphism of $\M_{3}(\C)$, then
\begin{equation}
Q(\cM)=Q^{\,\mr{a}}(\cM)=\frac{t^{2}}{81}\left[\frac{4}{9}\,\tr\,\cM\,\I_{3}\otimes
\I_{3}-2\sum\limits_{p=3,8}U\las{p}U^{\ast}\otimes
\tau_{I}(U)\tau_{I}(\las{p})\tau_{I}(U^{\ast})\right]\label{Q1}
\end{equation}
On the other hand, if $I$ defines anti - automorphism, then
\begin{equation}
Q(\cM)=Q^{\,\mr{aa}}(\cM)=\frac{t^{2}}{81}\left[\frac{4}{9}\,\tr\,\cM\,\I_{3}\otimes
\I_{3}+2\,\left(\sum\limits_{k}\las{k}\otimes\tau_{I}(\las{k})+\sum\limits_{p=3,8}U\las{p}U^{\ast}\otimes
\tau_{I}(U^{\ast})\tau_{I}(\las{p})\tau_{I}(U)\right)\right]\label{Q2}
\end{equation}
where $U\in \mr{SU(3)}$ is such that $\cM=R(U)\cM_{0}R(U)^{T}$.
\end{thm}
Proof:\\
Let us start with (\ref{QMI}) (for $d=3$) and consider
\begin{equation}
Y=\sum\limits_{j,k}Y_{jk}\las{j}\otimes\las{k}\label{Y}
\end{equation}
By a direct calculations, one  checks that the matrix (\ref{Y}) can be given by two equivalent representations
\begin{equation}
\begin{split}
Y=&-2\,\sum\limits_{p=3,8}U\las{p}U^{\ast}\otimes \tau_{I}(U\las{p}U^{\ast})+\frac{4}{3}\,\I_{3}\otimes \sum\limits_{p=3,8}
(\tau_{I}(U\las{p}U^{\ast}))^{2}-\frac{16}{9}\,
\I_{3}\otimes\I_{3}\\
&+\sum\limits_{j,k\neq 3,8}U\las{j}\las{k}U^{\ast}\otimes [\tau_{I}(U\las{j}U^{\ast})\tau_{I}(U\las{k}U^{\ast})-\tau_{I}(U\las{j}\las{k}U^{\ast})]
\end{split}\label{Y1}
\end{equation}
or
\begin{equation}
\begin{split}
Y=& 2\left(\sum\limits_{j}\las{j}\otimes\tau_{I}(\las{j})+\sum\limits_{p=3,8}U\las{p}U^{\ast}\otimes\tau_{I}(U\las{p}U^{\ast})\right)
+\frac{4}{3}\,\I_{3}\otimes \sum\limits_{p=3,8}(\tau_{I}(U\las{p}U^{\ast}))^{2}-\frac{16}{9}\,\I_{3}\otimes\I_{3}\\
&+\sum\limits_{j,k\neq 3,8}U\las{j}\las{k}U^{\ast}\otimes[\tau_{I}(U\las{j}U^{\ast})\tau_{I}(U\las{k}U^{\ast}))-\tau_{I}(U\las{k}\las{j}U^{\ast})]
\end{split}\label{Y2}
\end{equation}
Now if $\tau_{I}$ is an automorphism or anti- automorphism, then
\begin{equation*}
(\tau_{I}(U\las{3}U^{\ast}))^{2}+(\tau_{I}(U\las{8}U^{\ast}))^{2}=\tau_{I}((U\las{3}U^{\ast})^{2}+(U\las{8}U^{\ast})^{2})=
\tau_{I}(\las{3}^{2}+\las{8}^{2})=\frac{4}{3}\,\I_{3}
\end{equation*}
so the second terms in the formulas (\ref{Y1}) and (\ref{Y2}) vanish. Moreover, in the case of automorphisms the third  term in (\ref{Y1})
vanishes, whereas in the case of anti - automorphism the same happens in (\ref{Y2}). Since in both cases $X_{k}=0$, we obtain
the formulas (\ref{Q1}) and (\ref{Q2}).
\section{Generalization to qudits}
Detailed analysis of the qutrit case show that the simplification of the formula for $Q(\cM)$, that can lead to the exact analytical results
concerning trace norm quantum discord, is obtained for two classes of states.   The first class is defined by the correlation matrices
 belonging to the equivalence class $[\I_{8}]$. This class can be enlarged to contain all other states which are locally equivalent. The corresponding correlation matrices are in general non - diagonal and have the form
\begin{equation}
\cK^\mr{{a}}=t\,V,\quad V\in \mr{G}(3).\label{Ka}
\end{equation}
Notice that the mapping $\tau_{V}$ is an automorphism of $\M_{3}(\C)$. The second class is given by equivalence class $[I_{0}]$, where $\tau_{I_{0}}$
defines the transposition in $\M_{3}(\C)$. Again this class can be enlarged to contain the states with generally non - diagonal correlation matrices
\begin{equation}
\cK^{\mr{aa}}=t\,\cT,\quad   \cT= V_{1}I_{0}V_{2}^{T},\; V_{1},\,V_{2}\in \mr{G}(3)\label{Kaa}
\end{equation}
and in this case the mapping $\tau_{\cT}$ is an anti - automorphism.
\par
To extend the analysis to the case of arbitrary qudits, we consider generalizations of the above classes
 of states: the class $\fE^{\mr{a}}$ given by the correlation matrices as in (\ref{Ka}) but for $V\in \mr{G}(d)$ and the class $\fE^{\mr{aa}}$, where correlation matrices are defined as in (\ref{Kaa}), but for $V_{1},\, V_{2}\in \mr{G}(d)$ and the matrix $I_{0}$ represents the transposition in $\M_{d}(\C)$. $I_{0}$ is diagonal matrix with elements
\begin{equation}
\left(I_{0}\right)_{kk}=\frac{1}{2}\tr (\las{k}^{T}\las{k}),\quad k=1,\ldots,d^{2}-1\label{I0d}
\end{equation}
Since the mappings $\tau_{V}$ and $\tau_{\cT}$ are Jordan automorphisms of $\M_{d}(\C)$, all simplifications in the formula for $Q(\cM)$
obtained in the case $d=3$, are  valid, with proper modifications, also in general case. By a direct computations one obtains the following generalizations of the formulas (\ref{Q1}) and (\ref{Q2}):
\begin{equation}
Q^{\,\mr{a}}(\cM)=\frac{t^{2}}{d^{4}}\left[\left(\frac{2}{d}\right)^{2}d(d-1)\,\I_{d}\otimes
\I_{d}-2\sum\limits_{k=2}^{d}U\las{k^{2}-1}U^{\ast}\otimes
\tau_{V}(U)\tau_{\cT}(\las{k^{2}-1})\tau_{V}(U^{\ast})\right]
\end{equation}
\begin{equation}
Q^{\,\mr{aa}}(\cM)=\frac{t^{2}}{d^{4}}\left[\left(\frac{2}{d}\right)^{2}d(d-1)\,\I_{d}\otimes
\I_{d}+2\,\left((d-2)\sum\limits_{k}\las{k}\otimes\tau_{\cT}(\las{k})+\sum\limits_{k=2}^{d}U\las{k^{2}-1}U^{\ast}\otimes
\tau_{\cT}(U^{\ast})\tau_{\cT}(\las{k^{2}-1})\tau_{\cT}(U)\right)\right]
\end{equation}
\par
When $Q(\cM)$ equals to $Q(\cM)^{\,\mr{a}}$ or $Q(\cM)^{\,\mr{aa}}$, we are able to analytically find the spectrum of $Q(\cM)$. Observe that
\begin{equation}
Q^{\,\mr{a}}(\cM)=U\otimes \tau_{V}(U)\,Q^{\,\mr{a}}(\cM_{0})\,U^{\ast}\otimes \tau_{V}(U^{\ast})
\end{equation}
and
\begin{equation}
Q^{\,\mr{aa}}(\cM)=U\otimes \tau_{\cT}(U^{\ast})\,Q^{\,\mr{aa}}(\cM_{0})\,U^{\ast}\otimes \tau_{\cT}(U).
\end{equation}
So in both cases the spectrum of $Q(\cM)$ is the same as the
spectrum of $Q(\cM_{0})$ and to compute $D_{1}^{M}$ we need not to
minimize over all $\cM$. To find the spectrum  we take identity mapping in the case of
$Q^{\,\mr{a}}(\cM_{0})$ and  the transposition in the case of  $Q^{\,\mr{aa}}(\cM_{0})$. Let
\begin{equation}
L_{d}=\sum\limits_{k=2}^{d}\las{k^{2}-1}\otimes \las{k^{2}-1},\quad K_{d}=(d-2)\,\sum\limits_{j=1}^{d^{2}-1}\las{j}\otimes\las{j}^{T}
\end{equation}
Consider the spectrum of $L_{d}$ and $K_{d}+L_{d}$. To simplify the formulas, we introduce the notation: $\mu_{k}^{(\al_{k})}$ denotes
the eigenvalue $\mu_{k}$ with its multiplicity $\al_{k}$.
One can check that
\begin{equation*}
\sigma (L_{d})=\left\{\left(\frac{2(d-1)}{d}\right)^{(d)},\:\left(-\frac{2}{d}\right)^{(d(d-1))}\right\},
\end{equation*}
and
\begin{equation*}
\sigma
(K_{d}+L_{d})=\left\{\left(\frac{2(d^{2}(d-2)+1)}{d}\right)^{(1)},\:\left(-\frac{2(d-1)}{d}\right)^{(d(d-1))},\:\left(\frac{2}{d}\right)^{(d-1)}\right\}.
\end{equation*}
Since
\begin{equation*}
\sigma (Q^{\,\mr{a}}(\cM_{0}))=\frac{t^{2}}{d^{4}}\,\left[\left(\frac{2}{d}\right)^{2}d(d-1)-2\,\sigma (L_{d})\right]
\end{equation*}
and
\begin{equation*}
\sigma (Q^{\,\mr{aa}}(\cM_{0}))=\frac{t^{2}}{d^{4}}\,\left[\left(\frac{2}{d}\right)^{2}d(d-1)+2\,\sigma (K_{d}+L_{d})\right],
\end{equation*}
we obtain
\begin{equation}
\sigma (Q^{\,\mr{a}}(\cM_{0}))=\left\{\left(\frac{4t^{2}}{d^{4}}\right)^{(d(d-1))},\;
\left(0\right)^{(d)}\right\}
\end{equation}
and
\begin{equation}
\sigma
(Q^{\,\mr{aa}}(\cM_{0}))=\left\{\left(\frac{t^{2}(d-1)^{2}}{d^{4}}\right)^{(1)},\:\left(\frac{4t^{2}}{d^{4}}\right)^{(d-1)},\:
\left(0\right)^{(d(d-1))}\right\}
\end{equation}
Thus
\begin{equation}
\tr\,\sqrt{Q^{\,\mr{a}}(\cM_{0})}=\frac{2(d-1)}{d}|t|\quad\text{and}\quad \tr\,\sqrt{Q^{\,\mr{aa}}(\cM_{0})}=\frac{4(d-1)}{d^{2}}|t|
\end{equation}
and we arrive at the result
\begin{thm}
For all states $\ro\in \fE^{\mr{a}}$
\begin{equation}
D_{1}^{M}(\ro)=|t|,\quad -\frac{d}{2(d-1)}\leq t\leq \frac{d}{2(d+1)}
\end{equation}
Similarly, for all $\ro\in \fE^{\mr{aa}}$
\begin{equation}
D_{1}^{M}(\ro)=\frac{2}{d}\,|t|,\quad -\frac{d}{2(d^{2}-1)}\leq t\leq
\frac{d}{2}
\end{equation}
\end{thm}
\vskip 4mm\noindent
\begin{rem}
Notice that the class $\fE^{\mr{a}}$ contains the Werner states \cite{W}, which have a property that they are the only states satisfying
\begin{equation}
\ro = U\otimes U\,\ro\,U^{\ast}\otimes U^{\ast},\quad U\in \mr{SU(d)}\label{W}
\end{equation}
On the other hand, the class $\fE^{\mr{aa}}$ contains so called isotropic states \cite{HH}, which satisfy
\begin{equation}
\ro= U\otimes \conj{U}\,\ro\, U^{\ast}\otimes U^{T},\quad U\in \mr{SU(d)}\label{izo}
\end{equation}
Now assuming (\ref{W}) or (\ref{izo}), we can directly obtain that
\begin{equation*}
Q^{\mr{a}}(\cM)=U\otimes U\, Q^{\mr{a}}(\cM_{0})\,U^{\ast}\otimes U^{\ast}
\end{equation*}
or
\begin{equation*}
Q^{\mr{aa}}(\cM)=U\otimes\conj{U}\,Q^{\mr{aa}}(\cM_{0})\,U^{\ast}\otimes U^{T},
\end{equation*}
respectively. Our analysis shows the converse: the only states for which such relations are valid belong to the classes $\fE^{\mr{a}}$ or $\fE^{\mr{aa}}$.
\end{rem}
\section{Some examples of two - qudit states}
In this Section we apply the above results to some two - qudit
states. Description of the set of states of two - qudits is a highly
nontrivial problem so we restrict our analysis  to some specific
classes of states. We start with the class of  generalized Bell -
diagonal states of two qutids which form a simplex $\cW^{(d)}$
living in the $d^{2}$ - dimensional real linear space. The
construction of $\cW^{(d)}$ is as follows \cite{BHN1}. Let us fix
the basis $\ket{0},\, \ket{1},\, \ket{2},\,\ldots,\,\ket{d-1}$ for
one - qudit space $\C^{d}$. In the space of two qudits consider the
maximally entangled pure state
\begin{equation*}
\Psi_{00}=\frac{1}{\sqrt{d}}\,\sum\limits_{k=0}^{d-1}\ket{k}\otimes\ket{k}
\end{equation*}
Let $\D^{(d)}$ be the set of pairs of indices $(m,n)$, where $m,\,
n\,\in \Z_{d}$ i.e. the addition and multiplication of indices are
modulo $d$. For each $\al=(m,n)\in \D^{(d)}$, define the unitary
operator
\begin{equation*}
W_{\al}=W_{(m,n)}=\sum\limits_{k=0}^{d-1}e^{(2\pi
i/d)kn}\,\ket{k}\bra{k+m}
\end{equation*}
and
\begin{equation*}
\Psi_{\al}=(W_{\al}\otimes \I_{d})\Psi_{00},\quad
P_{\al}=\ket{\Psi_{\al}}\bra{\Psi_{\al}}
\end{equation*}
The class of generalized Bell - diagonal states contains all
mixtures of pure states $\Psi_{\al},\; \al\in \D^{(d)}$:
\begin{equation*}
\cW^{(d)}=\left\{\sum\limits_{\al\in \D^{(d)}}p_{\al}P_{\al}\,:\,
p_{\al}\geq 0,\; \sum\limits_{\al}p_{\al}=1\right\}
\end{equation*}
Any state $\ro\in \cW^{(d)}$ is locally maximally mixed and the
corresponding correlation matrix belong to the simplex
$\K_{\cW}^{(d)}\subset \K^{(d)}$ spanned by matrices $\cK_{\al}$,
which are correlation matrices of $\Psi_{\al}$. One can check that
\begin{equation}
\cK_{00}=\frac{d}{2}\,I_{0},\label{K00}
\end{equation}
where $I_{0}$ is given by (\ref{I0d}) and for any other $\al\in
\D^{(d)}$
\begin{equation*}
\cK_{\al}=\cK_{00}\,V_{\al}^{T},\quad  V_{\al}=R(W_{\al})
\end{equation*}
so the set $\K_{\cW}^{(d)}$ is given by convex combinations of the
matrices which are orthogonal up to the multiplicative constant.
Notice that the maximally mixed state
$\ro_{\infty}=\frac{1}{d^{2}}\I_{d^{2}}$ lies at the center of
$\cW^{(d)}$
\begin{equation*}
\ro_{\infty}=\frac{1}{d^{2}}\,\sum\limits_{\al\in \D^{(d)}}P_{\al}
\end{equation*}
and the corresponding correlation matrix is the zero matrix.
\par
Now we consider quantum discord of some states from the simplex
$\cW^{(d)}$. We start with the pure states $P_{\al}$. Since all such
states are equivalent it is enough to take one of them, for example
$P_{00}$. The state $P_{00}$  has the correlation matrix
(\ref{K00}), so  it belongs to the class $\fE^{\mr{aa}}$ and by
Theorem 6,
 $D_{1}^{M}(P_{00})=1$. Thus for all $\al\in \D^{(d)}$
\begin{equation*}
D_{1}^{M}(P_{\al})=D_{2}^{M}(P_{\al})=1
\end{equation*}
Take the convex combination
\begin{equation}
\ro_{\mr{iso},\al}=(1-p)\ro_{\infty}+pP_{\al}.\label{iso}
\end{equation}
For a fixed value of $p$, such states are locally equivalent and
\begin{equation*}
\cK_{\mr{iso},\al}=p\,\cK_{\al}
\end{equation*}
so $\ro_{\mr{iso},\al}$ are isotropic and
\begin{equation*}
D_{1}^{M}(\ro_{\mr{iso},\al})=\sqrt{D_{2}^{M}(\ro_{\mr{iso},\al})}=p,\quad p\in [0,1]
\end{equation*}
The explicit construction of density matrices (\ref{iso}) is not
difficult, but instead of it we can easily find the corresponding
correlation matrices. To give an example beyond the qutrit case, we
take  $d=4$. One can check that
\begin{equation}
\cK_{\mr{iso},(00)}=2p\;\mr{diag}\,(1,-1,1,1,-1,1,-1,1,1,-1,1,-1,1,-1,1).
\end{equation}
\par
More detailed analysis of the states from the simplex $\cW^{(d)}$ is
technically very involved and we restrict it to the case of
$\cW^{(3)}$. As it was shown in \cite{BHN}, there are some
equivalences inside $\cW^{(3)}$ which can help in computations. It
turns out that local operations on elements of $\cW^{(3)}$ can be
identified with affine transformations of the set $\D^{(3)}$ and all
subsets of $\D^{(3)}$ can be classified with respect to this local
equivalence relation. In particular there is one class of single
point, one of two points and two classes of three points. The subset
$\ell\subset \D^{(3)}$ of the form
\begin{equation*}
\ell=\{(j,k),\, (j+n,k+n),\, (j+2n,k+2n)\}
\end{equation*}
is called a line in $\D^{(3)}$. It can be shown that all lines are
equivalent. All other sets of three points form another equivalence
class. To the equivalence classes in $\D^{(3)}$ correspond
equivalence classes in the simplex $\cW^{(3)}$. Single points $\al$
define pure states $P_{\al}$ which are equivalent by construction.
Each pair $\{\al, \beta\}$ gives a mixture
\begin{equation}
\ro_{\{\al,\beta\}}=p_{\al}P_{\al}+p_{\beta}P_{\beta}\label{roalbe}
\end{equation}
and all such states (for fixed $p_{\al},\, p_{\beta}$) are locally
equivalent. For any line $\ell$ the states
\begin{equation}
\ro_{\ell}=\sum\limits_{\al\in \ell}p_{\al}P_{\al}\label{roline}
\end{equation}
are locally equivalent, another class forms the states
\begin{equation*}
\ro_{\{\al,\,\beta,\,
\gamma\}}=p_{\al}P_{\al}+p_{\beta}P_{\beta}+p_{\gamma}P_{\gamma}
\end{equation*}
for $\al,\, \beta,\, \gamma$ not lying on any line.
\par
For the mixture (\ref{roalbe}) we are not able to find analytic
expression for quantum discord, so  we look for a lower bound, given
by Theorem 1. In this case we have
\begin{equation*}
\cK_{\{\al,\beta\}}=p_{\al}\,\cK_{\al}+p_{\beta}\,\cK_{\beta}
\end{equation*}
and it is enough to take particular points, for example
$\al=(0,0),\; \beta=(2,2)$. In this case we have
\begin{equation*}
\cK_{\{\al,\beta\}}=
\begin{pmatrix}
 \frac{3}{2}p_{\al} & 0 & 0 & -\frac{3}{4}p_{\beta} & \frac{3 \sqrt{3}}{4}p_{\beta} & 0 & 0 & 0
 \\[2mm]
 0 & -\frac{3}{2}p_{\al} & 0 & -\frac{3 \sqrt{3}}{4}p_{\beta} & -\frac{3 }{4}p_{\beta} & 0 & 0 & 0
 \\[2mm]
 0 & 0 & \frac{3}{4} (2p_{\al}-p_{\beta}) & 0 & 0 & 0 & 0 & -\frac{3 \sqrt{3}}{4}p_{\beta}
 \\[2mm]
 0 & 0 & 0 & \frac{3}{2}p_{\al} & 0 & -\frac{3}{4}p_{\beta} & -\frac{3 \sqrt{3}}{4}p_{\beta} & 0
 \\[2mm]
 0 & 0 & 0 & 0 & -\frac{3}{2}p_{\al} & \frac{3 \sqrt{3}}{4}p_{\beta} & -\frac{3}{4}p_{\beta} & 0
 \\[2mm]
 -\frac{3}{4}p_{\beta} & -\frac{3 \sqrt{3}}{4}p_{\beta} & 0 & 0 & 0 & \frac{3}{2}p_{\al} & 0 & 0
 \\[2mm]
 -\frac{3 \sqrt{3}}{4}p_{\beta} & \frac{3}{4}p_{\beta} & 0 & 0 & 0 & 0 & -\frac{3}{2}p_{\al} & 0
 \\[2mm]
 0 & 0 & \frac{3 \sqrt{3}}{4}p_{\beta} & 0 & 0 & 0 & 0 & \frac{3}{4} (2p_{\al}-p_{\beta})
\end{pmatrix}
\end{equation*}
and one can check that the matrix
$\cK_{\{\al,\beta\}}\cK_{\{\al,\beta\}}^{T}$ has the eigenvalues:
$\frac{9}{4}$ with multiplicity $2$ and
$\frac{9}{4}(1-3p_{\al}p_{\beta})$ with multiplicity $6$, so
\begin{equation*}
\Xi(\cK_{\{\al,\beta\}})=\frac{27}{2}(1-3p_{\al}p_{\beta})
\end{equation*}
Applying Theorem 2 we obtain that
\begin{equation*}
D_{2}^{M}(\ro_{\{\al,\beta\}})\geq 1-3p_{\al}p_{\beta},\quad
D_{1}^{M}(\ro_{\{\al,\beta\}})\geq
\sqrt{\frac{3}{8}(1-3p_{\al}p_{\beta})}
\end{equation*}
It is interesting to note that this bound can be expressed in term
of negativity of the state  \cite{VW}
\begin{equation*}
N(\ro)=\frac{1}{2}(||\ro^{PT}||_{1}-1)
\end{equation*}
where $\ro^{PT}$ denotes partial transposition of  $\ro$.
As it was shown in Ref. \cite{DJ}
\begin{equation*}
N(\ro_{\{\al,\beta\}})=\sqrt{1-3p_{\al}p_{\beta}}
\end{equation*}
so
\begin{equation*}
D_{2}^{M}(\ro_{\{\al,\beta\}})\geq N(\ro_{\{\al,\beta\}})^{2},\quad
D_{1}^{M}(\ro_{\{\al,\beta\}})\geq
\sqrt{\frac{3}{8}}\,N(\ro_{\{\al,\beta\}})
\end{equation*}
Notice that the states $\ro_{\{\al,\beta\}}$ have always non - zero
quantum discord and the minimal value of lower bound is attained for
a symmetric mixture of pure states.
\par
Consider now the mixture $\ro_{\ell}$ given by any line  $\ell \in
\D$. Since all such mixtures are locally equivalent, take
$\ell=\{(0,0),\, (1,1),\, (2,2)\}$, then
$\cK_{\ell}=p_{\al}\cK_{\al}+p_{\beta}\cK_{\beta}+p_{\gamma}\cK_{\gamma}$
is given by
\begin{equation*}
\cK_{\ell}=
\begin{pmatrix}
 \frac{3}{2} p_{\alpha } & 0 & 0 & -\frac{3}{4} p_{\gamma } & \frac{3 \sqrt{3}}{4} p_{\gamma } &
 -\frac{3}{4} p_{\beta } &
   \frac{3 \sqrt{3}}{4} p_{\beta } & 0 \\[2mm]
 0 & -\frac{3}{2} p_{\alpha } & 0 & -\frac{3 \sqrt{3}}{4} p_{\gamma } & -\frac{3}{4} p_{\gamma } &
 \frac{3 \sqrt{3}}{4} p_{\beta} & \frac{3}{4} p_{\beta } & 0 \\[2mm]
 0 & 0 & \frac{3}{4} \left(2 p_{\alpha }-p_{\beta }-p_{\gamma }\right) & 0 & 0 & 0 & 0 &
 \frac{3 \sqrt{3}}{4} \left(p_{\beta}-p_{\gamma }\right) \\[2mm]
 -\frac{3}{4} p_{\beta } & \frac{3 \sqrt{3}}{4} p_{\beta } & 0 & \frac{3}{2} p_{\alpha } & 0 &
 -\frac{3}{4} p_{\gamma } &
   -\frac{3 \sqrt{3}}{4} p_{\gamma } & 0 \\[2mm]
 -\frac{3 \sqrt{3}}{4} p_{\beta } & -\frac{3}{4} p_{\beta } & 0 & 0 & -\frac{3}{2} p_{\alpha } &
 \frac{3 \sqrt{3}}{4} p_{\gamma} & -\frac{3}{4} p_{\gamma } & 0
 \\[2mm]
 -\frac{3}{4} p_{\gamma } & -\frac{3 \sqrt{3}}{4} p_{\gamma } & 0 & -\frac{3}{4} p_{\beta } &
 -\frac{3 \sqrt{3}}{4} p_{\beta} & \frac{3}{2} p_{\alpha } & 0 & 0
 \\[2mm]
 -\frac{3 \sqrt{3}}{4} p_{\gamma } & \frac{3}{4} p_{\gamma } & 0 & \frac{3 \sqrt{3}}{4} p_{\beta } &
 -\frac{3}{4} p_{\beta }
   & 0 & -\frac{3}{2} p_{\alpha } & 0 \\[2mm]
 0 & 0 & -\frac{3 \sqrt{3}}{4} \left(p_{\beta }-p_{\gamma }\right) & 0 & 0 & 0 & 0 & \frac{3}{4}
 \left(2 p_{\alpha}-p_{\beta }-p_{\gamma }\right)
\end{pmatrix}
\end{equation*}
The matrix $\cK_{\ell}\cK_{\ell}^{T}$ has the following eigenvalues:
$\frac{9}{4}$ with multiplicity $2$ and
 $\frac{9}{8}\sum\limits_{\al\neq \beta\in
\ell}(p_{\al}-p_{\beta})^{2}$ with multiplicity $6$. So
\begin{equation*}
\Xi(\cK_{\ell})=\frac{27}{4}\sum\limits_{\al\neq \beta\in
\ell}(p_{\al}-p_{\beta})^{2}
\end{equation*}
Also in this case the lower bound is given by negativity of the
state, since \cite{DJ}
\begin{equation*}
N(\ro_{\ell})=\sqrt{\frac{1}{2}\sum\limits_{\al\neq \beta\in
\ell}(p_{\al}-p_{\beta})^{2}}
\end{equation*}
So, similarly as in the case of $\ro_{\{\al,\beta\}}$ we have
\begin{equation*}
D_{2}^{M}(\ro_{\ell})\geq N(\ro_{\ell})^{2},\quad
D_{1}^{M}(\ro_{\ell})\geq \sqrt{\frac{3}{8}}\,N(\ro_{\ell})
\end{equation*}
Notice that this time  for symmetric mixture of pure states the
lower bound is equal to zero.
\par
As it was proved in Theorem 6, the exact value of $D_{1}^{M}$ can be
also obtained for a class $\fE^{\mr{a}}$ containing Werner states.
Qutrit Werner states do not belong to the simplex $\cW^{(3)}$, but
are examples of the states with diagonal orthogonal correlation
matrices i.e. matrices $I$ satisfying $I^{2}=\I_{8}$ (see Section
IV). Let $\cJ_{8}$ denotes the set of real diagonal $8\times 8$
matrices $I$ satisfying $I^{2}=\I_{8}$. For any $I\in \cJ_{8}$ the
formula
\begin{equation}
\ro=\frac{1}{9}\,\left(\I_{3}\otimes
\I_{3}+t\,\sum\limits_{k=1}^{8}\ip{Ie_{k}}{\la}\otimes
\ip{e_{k}}{\la}\right), \label{qstate}
\end{equation}
defines a state of two qutrits for a suitable range of values of the
parameter $t$ (depending on the matrix $I$). Let us denote the set
of such states by $\fE_{\cJ}$. For all $ \ro \in\fE_{\cJ}$ , we
have
\begin{equation}
D^{M}_{2}(\ro)=\frac{4}{9}\,t^{2},\quad D^{M}_{1}(\ro)\geq \frac{1}{\sqrt{6}}\,|t| \label{DI}
\end{equation}
but the entanglement properties of these states  depend on the
choice of the matrix $I$.  There are $2^{8}$ of such
states and the set $\fE_{\cJ}$ can be divided into $16$ isospectral classes
$\fE_{k}$ of states with the same spectrum. In fact there are only
$8$ independent classes, since the remaining classes can be obtained
by simple reparametrization $ t\to -t$. It turns out that
\begin{equation*}
\fE_{k}=\bigcup\limits_{p}\fE_{k,p}
\end{equation*}
where the  states from $\fE_{k,p}$ are given by matrices  $I\in [I_{k,p}]$ and  the class $[I_{k,p}]$
is defined as
\begin{equation*}
[I_{k,p}]=\{I_{k,p},\, I_{k,p}V_{1},\, I_{k,p}V_{2},\,I_{k,p}V_{3}\},
\end{equation*}
for some $I_{k,p}\in \cJ_{8}$ and  $V_{1}=R(W_{1}),\, V_{2}=R(W_{2}),\, V_{3}=R(W_{3})$, where $W_{1},\, W_{2},\, W_{3}$ are given by (\ref{W123}).
It is obvious that the states in each $\fE_{k,p}$ are locally
equivalent and in this sense the class $[I_{k,p}]$ is an equivalence
class. In the Appendix C we list all the isospectral classes of two - qutrit states and discuss entanglement properties of states from $\fE_{\cJ}$. It follows that the Werner states belong to the class defined by $[\I_{8}]$, which is a subclass of the isospectral class $\fE_{3}$  containing $16$ density matrices with eigenvalues
$\frac{1}{27}\,(3-8t)$  with multipliticity $2$ and $\frac{1}{27}\,(3+4t)$ with multipliticity $6$. Moreover  $t\in [-3/4,3/8]$. There are three other   subclasses of $\fE_{3}$ containing equivalent states, but the states from different subclasses are not locally equivalent. The subclasses are defined by matrices $I_{3,2},\, I_{3,3}$ and $I_{3,4}$ (see Appendix C for the notation and all details). Now we can apply the general result to the qutrit Werner states. By Theorem 6 for all states in the class $\fE_{3,1}$ we have
\begin{equation*}
D_{1}^{M}(\ro)=\frac{2}{3}\,|t|,\quad -\frac{3}{4}\leq t\leq \frac{3}{8}
\end{equation*}
Notice that  there is a small common part of the set $\fE_{\cJ}$ and the simplex $\cW$, since  the set $\fE_{\cJ}$ contains  the isotropic states which belong to the subclass defined by the matrix $I_{0}$.

\section{Conclusions}
In the present paper we derived the explicit form of the
measurement-induced quantum discord for arbitrary $d$-level system.
The complexity of the problem is such that it was  possible only for
states within the  selected classes of correlation matrices.  Type
of difficulties, not only degree, changes between the $d=2$ case and
$d \geq 3 $. The qubit systems are very special and their simpler to
study properties, are strictly related with the geometry of the state
space, which in this case is a geometry of the unit ball in
$\R^{3}$. Moreover, the group $\mr{SU(2)}$ is homomorphic with the
orthogonal group $\mr{SO(3)}$. For higher dimensions, the geometry
of one-qudit quantum state space is very rich  and is known to some
extend only for qutrits \cite{GSSS}, but not known for generic case.
Concerning the properties of adjoint representation of $\mr{SU(d)}$
it is known that when $d\geq 3$, $\mr{G(d)}$ is isomorphic merely to
a relatively small subgroup of $\mr{SO(d^{2}-1)}$.
\par
During the process of computation of quantum discord, the main
obstacle is to find analytically the minimum of the square of state
disturbance $Q(\cM)$ over all projectors $\cM$ in $\R^{d^{2}-1}$
coming from local measurements. Such kind of minimization problem
can be explicitly solved when we consider all orthogonal projectors
on $d(d-1)$ dimensional subspaces of $\R^{d^{2}-1}$ (Lemma 2). Since
$\cM=V\cM_{0}V^{T}$, where $\cM_{0}$ corresponds to the canonical
von Neumann measurement and $V\in \mr{G}(d)$, only in the case of
qubits where $\mr{G}(2)=\mr{SO(3)}$, the projectors $\cM$ run over
the set of all projectors with a fixed dimension of its range and
one is able to find the analytic formula for  Hilbert - Schmidt
geometric discord. When $d\geq 3$,  this set is a proper subset of
all such orthogonal projectors. From this perspective  qubits are
essentially different then qutrits, which in turn are more similar
to  the rest of higher dimensional qudits and for them even in the
case of the Hilbert-Schmidt distance  we are able to find only a
lower bound for geometric discord.
\par
In our study we focused on locally maximally mixed (LMM) states i.e.
such states that the restrictions to subsystems are maximally mixed.
In the coherence vector representation LMM state is fully described
by the correlation matrix $\cK$. Again the cases of qubits and
qudits differ significantly. Since in the case of qubits we can use
full orthogonal group of transformation to diagonalize correlation
matrix, any two-qubit LMM state is localy equivalent to the state
with diagonal $\cK$. For higher dimensional qudits it is generally
not true (not all $\mr{SO(d^{2}-1)}$ transformations are at our
disposal) and we cannot restrict the analysis to the diagonal case.
For any non-diagonal correlation matrix we were able to find a lower
bound on measurement-induced qudit discord in terms of sigular
values of the matrix $\cK$. In particular we show that when
$\mr{rank}\, \cK\geq d$, the state has non-zero quantum discord.
However, for suitably chosen  $\cK$ we obtain more. Namely, when the
correlation matrix is proportional to some orthogonal matrix on
$\R^{d^{2}-1}$ it is possible to compute the exact value of
Hilbert-Schmidt geometric discord and find the lower bound on its
trace-norm counterpart. To obtain the exact value of trace-norm
discord we must find the spectrum of the  square
 of  disturbance of the state induced by any local projective measurement. In general this problem is
hard (or even non-tractable \cite{H}), so we look for further
simplifications by exploiting underlying geometry of qudits, in
particular the interesting properties of the $\star$-product in
$\R^{d^{2}-1}$.
 We performed a detailed analysis of this problem in the case of qutrits and
generalized it to arbitrary qudits. The main result shows that if
the orthogonal matrix on $\R^{d^{2}-1}$ defines (via the
parametrization in terms of generators of the Lie algebra
$\mr{su(d)}$) a Jordan automorphism of the algebra of $d\times d$
matrices, the spectrum of the square of state disturbance does not
depend on  local measurements, so to compute the  quantum discord in
this case, the minimization procedure is not necessary. Any Jordan
automorphism of the matrix algebra is either an automorphism or
anti-automorphism and in our case this gives two equivalence classes
(with respect to adjoint representation of $\mr{SU(d)}$) of
correlation matrices: one containing identity matrix and the other
containing  the orthogonal matrix implementing the transposition.
The corresponding classes of states are interesting for many
reasons. In particular, the first class contains so called Werner
states  and the second contains isotropic states. In both cases we
find the spectrum of the square  of the state disturbance and
compute analytically the value of trace-norm measurement induced
geometric discord. Finally, we applied the obtained results to some
specific classes of two-qudit states. For
Bell-diagonal two-qutrit states and the family of states
with diagonal orthogonal correlation matrices we have studied the
relation between the measure of discord and the measure of
entanglement, given by negativity.
 \par
Our research contributes to the ongoing discussion on the best
choice of preferred measure of quantum discord. We have shown that
the measurement-induced quantum geometric discord based on the trace
norm  can be effectively computed despite that it is perceived as
less easy to handle then the one defined by the Hilbert - Schmidt
distance. What is important it belongs to the set of bona fide
measures of quantum correlations and definitely deserves further
study.



\section*{Appendix A}
\noindent\textit{Proof of Lemma 4}:\\[4mm]
Let $I$ be a diagonal $8 \times 8$ matrix satisfying $I^{2}=\I_{8}$. Define $S\subset \{1,2,\ldots,8\}$ by $
S=\{ m\,:\, I_{mm}=-1\}$ (we exclude the case $I=\I_{8}$) and let $S^{c}=\{1,2,\ldots,8\}\setminus S$. If the condition (\ref{IMIstar})
is satisfied, then $\tr (\Delta_{l}I\cM I)=0$ for all $l=1,\ldots,8$. Notice that
\begin{equation*}
\tr (I\Delta I\cM)=4\, \sum\limits_{p=3,8}\sum\limits_{m\in S,\, q\in S^{c}}\Delta_{l, mq}V_{mp}V_{qp}
\end{equation*}
and
\begin{equation*}
\sum\limits_{l}\Delta_{m,lq}\las{l}=\Delta_{m}\las{q}=\frac{1}{2}\,[\las{m},\las{q}]_{+}-\frac{2}{3}\,\delta_{mq}\,\I_{3}
\end{equation*}
so the condition (\ref{IMIstar}) can be written as
\begin{equation*}
\sum\limits_{p=3,8}\left(\sum\limits_{m\in S}V_{mp}\las{m}\cdot\sum\limits_{q\in S^{c}}V_{qp}\las{q}+\sum\limits_{q\in S^{c}}V_{qp}\las{q}
\cdot\sum\limits_{m\in S}V_{mp}\las{m}\right)=0
\end{equation*}
Since
\begin{equation*}
\sum\limits_{i\in S}V_{ij}\las{i}=-\frac{1}{2}\left(\tau_{I}(U\las{j}U^{\ast})-U\las{j}U^{\ast}\right),\quad
\sum\limits_{i\in S^{c}}V_{ij}\las{i}=\frac{1}{2}\left(\tau_{I}(U\las{j}U^{\ast})+U\las{j}U^{\ast}\right)
\end{equation*}
the condition (\ref{IMIstar}) gives
\begin{equation*}
\sum\limits_{p=3,8}(\tau_{I}(U\las{p}U^{\ast}))^{2}=\sum\limits_{p=3,8}(U\las{p}U^{\ast})^{2}=
U\,\left(\sum\limits_{p=3,8}\las{p}^{2}\right)\,U^{\ast}=\frac{4}{3}\,\I_{3}
\end{equation*}
\vskip 4mm\noindent
\textit{Proof of Lemma 5}:\\[4mm]
Consider $I=\mr{diag}\,(\ee_{1},\ldots,\ee_{7},1),\; \ee_{k}\in \{-1,1\}$. Let $U\in \mr{SU(3)}$ and the last row of $U$ is given by complex numbers $(a,b,c)$
where $|a|^{2}+|b|^{2}+|c|^{2}+1$. Denote $a_{R}=\mr{Re}\,a,\; a_{J}=\mr{Im}\,a$ and similarly for $b$ and $c$. Take the mapping $(a,b,c)\to (a^{\prime},b^{\prime},c^{\prime})$ where
\begin{equation*}
a^{\prime}_{R}=\ee_{1}a_{R},\,a^{\prime}_{J}=\ee_{2}a_{J};\quad b^{\prime}_{R}=b_{R},\, b^{\prime}_{J}=\ee_{1}\ee_{2}b_{j};\quad
c^{\prime}_{R}=\ee_{2}\ee_{5}c_{R},\, c^{\prime}_{J}=\ee_{1}\ee_{5}c_{J}
\end{equation*}
Let transform any row of $U$ in this way and denote the resulting matrix by $\widetilde{U}$ and  define $W=(\mr{det}\,\widetilde{U})^{-1}\,\widetilde{U}$. One checks that $W\in \mr{SU(3)}$. Define also
\begin{equation*}
I_{+}=\mr{diag}\,(\ee_{1},\ee_{2},1,\ee_{1}\ee_{2}\ee_{5},\ee_{5},\ee_{2}\ee_{5},\ee_{1}\ee_{5},1)
\end{equation*}
and
\begin{equation*}
J=\mr{diag}\,(1,1,\ee_{3},\ee_{1}\ee_{2}\ee_{4}\ee_{5},1,\ee_{2}\ee_{5}\ee_{6},\ee_{1}\ee_{5}\ee_{7},1)
\end{equation*}
then $I=I_{+}J$ and $\tau_{I_{+}}$ is positive since
\begin{equation*}
\tau_{I_{+}}(UP_{3}U^{\ast})=WP_{3}W^{\ast},\quad P_{3}=\I_{3}-\las{3}^{2}
\end{equation*}
Moreover, $\tau_{I_{+}}$ satisfies the condition (\ref{la3la8}). On the other hand, one can show that
$\tau_{J}$ is not positive and this mapping violates the condition (\ref{la3la8}). It implies that
$\tau_{I}=\tau_{I_{+}}\circ \tau_{J}$ is not positive, unless $J=\I_{8}$. Similarly one can show that $\tau_{I}$
can not satisfy the condition (\ref{la3la8}).
\section*{Appendix B}
Matrices $V_{\al}$:
\begin{equation*}
V_{01}=
\begin{pmatrix}
 -\frac{1}{2} & \frac{\sqrt{3}}{2} & 0 & 0 & 0 & 0 & 0 & 0 \\[2mm]
 -\frac{\sqrt{3}}{2} & -\frac{1}{2} & 0 & 0 & 0 & 0 & 0 & 0 \\[2mm]
 0 & 0 & 1 & 0 & 0 & 0 & 0 & 0 \\[2mm]
 0 & 0 & 0 & -\frac{1}{2} & -\frac{\sqrt{3}}{2} & 0 & 0 & 0 \\[2mm]
 0 & 0 & 0 & \frac{\sqrt{3}}{2} & -\frac{1}{2} & 0 & 0 & 0 \\[2mm]
 0 & 0 & 0 & 0 & 0 & -\frac{1}{2} & \frac{\sqrt{3}}{2} & 0 \\[2mm]
 0 & 0 & 0 & 0 & 0 & -\frac{\sqrt{3}}{2} & -\frac{1}{2} & 0 \\[2mm]
 0 & 0 & 0 & 0 & 0 & 0 & 0 & 1
\end{pmatrix}
,\quad V_{02}=
\begin{pmatrix}
 0 & 0 & 0 & 0 & 0 & 1 & 0 & 0 \\[2mm]
 0 & 0 & 0 & 0 & 0 & 0 & 1 & 0 \\[2mm]
 0 & 0 & -\frac{1}{2} & 0 & 0 & 0 & 0 & \frac{\sqrt{3}}{2} \\[2mm]
 1 & 0 & 0 & 0 & 0 & 0 & 0 & 0 \\[2mm]
 0 & -1 & 0 & 0 & 0 & 0 & 0 & 0 \\[2mm]
 0 & 0 & 0 & 1 & 0 & 0 & 0 & 0 \\[2mm]
 0 & 0 & 0 & 0 & -1 & 0 & 0 & 0 \\[2mm]
 0 & 0 & -\frac{\sqrt{3}}{2} & 0 & 0 & 0 & 0 & -\frac{1}{2}
\end{pmatrix}
\end{equation*}
\begin{equation*}
V_{10}=
\begin{pmatrix}
 -\frac{1}{2} & \frac{\sqrt{3}}{2} & 0 & 0 & 0 & 0 & 0 & 0 \\[2mm]
 -\frac{\sqrt{3}}{2} & -\frac{1}{2} & 0 & 0 & 0 & 0 & 0 & 0 \\[2mm]
 0 & 0 & 1 & 0 & 0 & 0 & 0 & 0 \\[2mm]
 0 & 0 & 0 & -\frac{1}{2} & -\frac{\sqrt{3}}{2} & 0 & 0 & 0 \\[2mm]
 0 & 0 & 0 & \frac{\sqrt{3}}{2} & -\frac{1}{2} & 0 & 0 & 0 \\[2mm]
 0 & 0 & 0 & 0 & 0 & -\frac{1}{2} & \frac{\sqrt{3}}{2} & 0 \\[2mm]
 0 & 0 & 0 & 0 & 0 & -\frac{\sqrt{3}}{2} & -\frac{1}{2} & 0 \\[2mm]
 0 & 0 & 0 & 0 & 0 & 0 & 0 & 1
\end{pmatrix}
,\quad V_{20}=
\begin{pmatrix}
 -\frac{1}{2} & -\frac{\sqrt{3}}{2} & 0 & 0 & 0 & 0 & 0 & 0 \\[2mm]
 \frac{\sqrt{3}}{2} & -\frac{1}{2} & 0 & 0 & 0 & 0 & 0 & 0 \\[2mm]
 0 & 0 & 1 & 0 & 0 & 0 & 0 & 0 \\[2mm]
 0 & 0 & 0 & -\frac{1}{2} & \frac{\sqrt{3}}{2} & 0 & 0 & 0 \\[2mm]
 0 & 0 & 0 & -\frac{\sqrt{3}}{2} & -\frac{1}{2} & 0 & 0 & 0 \\[2mm]
 0 & 0 & 0 & 0 & 0 & -\frac{1}{2} & -\frac{\sqrt{3}}{2} & 0 \\[2mm]
 0 & 0 & 0 & 0 & 0 & \frac{\sqrt{3}}{2} & -\frac{1}{2} & 0 \\[2mm]
 0 & 0 & 0 & 0 & 0 & 0 & 0 & 1
\end{pmatrix}
\end{equation*}
\begin{equation*}
V_{11}=
\begin{pmatrix}
 0 & 0 & 0 & -\frac{1}{2} & -\frac{\sqrt{3}}{2} & 0 & 0 & 0 \\[2mm]
 0 & 0 & 0 & -\frac{\sqrt{3}}{2} & \frac{1}{2} & 0 & 0 & 0 \\[2mm]
 0 & 0 & -\frac{1}{2} & 0 & 0 & 0 & 0 & -\frac{\sqrt{3}}{2} \\[2mm]
 0 & 0 & 0 & 0 & 0 & -\frac{1}{2} & \frac{\sqrt{3}}{2} & 0 \\[2mm]
 0 & 0 & 0 & 0 & 0 & \frac{\sqrt{3}}{2} & \frac{1}{2} & 0 \\[2mm]
 -\frac{1}{2} & \frac{\sqrt{3}}{2} & 0 & 0 & 0 & 0 & 0 & 0 \\[2mm]
 -\frac{\sqrt{3}}{2} & -\frac{1}{2} & 0 & 0 & 0 & 0 & 0 & 0 \\[2mm]
 0 & 0 & \frac{\sqrt{3}}{2} & 0 & 0 & 0 & 0 & -\frac{1}{2}
\end{pmatrix}
,\quad V_{12}=
\begin{pmatrix}
 0 & 0 & 0 & 0 & 0 & -\frac{1}{2} & \frac{\sqrt{3}}{2} & 0 \\[2mm]
 0 & 0 & 0 & 0 & 0 & -\frac{\sqrt{3}}{2} & -\frac{1}{2} & 0 \\[2mm]
 0 & 0 & -\frac{1}{2} & 0 & 0 & 0 & 0 & \frac{\sqrt{3}}{2} \\[2mm]
 -\frac{1}{2} & \frac{\sqrt{3}}{2} & 0 & 0 & 0 & 0 & 0 & 0 \\[2mm]
 \frac{\sqrt{3}}{2} & \frac{1}{2} & 0 & 0 & 0 & 0 & 0 & 0 \\[2mm]
 0 & 0 & 0 & -\frac{1}{2} & -\frac{\sqrt{3}}{2} & 0 & 0 & 0 \\[2mm]
 0 & 0 & 0 & -\frac{\sqrt{3}}{2} & \frac{1}{2} & 0 & 0 & 0 \\[2mm]
 0 & 0 & -\frac{\sqrt{3}}{2} & 0 & 0 & 0 & 0 & -\frac{1}{2}
\end{pmatrix}
\end{equation*}
\begin{equation*}
V_{21}=
\begin{pmatrix}
 0 & 0 & 0 & -\frac{1}{2} & \frac{\sqrt{3}}{2} & 0 & 0 & 0 \\[2mm]
 0 & 0 & 0 & \frac{\sqrt{3}}{2} & \frac{1}{2} & 0 & 0 & 0 \\[2mm]
 0 & 0 & -\frac{1}{2} & 0 & 0 & 0 & 0 & -\frac{\sqrt{3}}{2} \\[2mm]
 0 & 0 & 0 & 0 & 0 & -\frac{1}{2} & -\frac{\sqrt{3}}{2} & 0 \\[2mm]
 0 & 0 & 0 & 0 & 0 & -\frac{\sqrt{3}}{2} & \frac{1}{2} & 0 \\[2mm]
 -\frac{1}{2} & -\frac{\sqrt{3}}{2} & 0 & 0 & 0 & 0 & 0 & 0 \\[2mm]
 \frac{\sqrt{3}}{2} & -\frac{1}{2} & 0 & 0 & 0 & 0 & 0 & 0 \\[2mm]
 0 & 0 & \frac{\sqrt{3}}{2} & 0 & 0 & 0 & 0 & -\frac{1}{2}
\end{pmatrix}
,\quad V_{22}=
\begin{pmatrix}
 0 & 0 & 0 & 0 & 0 & -\frac{1}{2} & -\frac{\sqrt{3}}{2} & 0 \\[2mm]
 0 & 0 & 0 & 0 & 0 & \frac{\sqrt{3}}{2} & -\frac{1}{2} & 0 \\[2mm]
 0 & 0 & -\frac{1}{2} & 0 & 0 & 0 & 0 & \frac{\sqrt{3}}{2} \\[2mm]
 -\frac{1}{2} & -\frac{\sqrt{3}}{2} & 0 & 0 & 0 & 0 & 0 & 0 \\[2mm]
 -\frac{\sqrt{3}}{2} & \frac{1}{2} & 0 & 0 & 0 & 0 & 0 & 0 \\[2mm]
 0 & 0 & 0 & -\frac{1}{2} & \frac{\sqrt{3}}{2} & 0 & 0 & 0 \\[2mm]
 0 & 0 & 0 & \frac{\sqrt{3}}{2} & \frac{1}{2} & 0 & 0 & 0 \\[2mm]
 0 & 0 & -\frac{\sqrt{3}}{2} & 0 & 0 & 0 & 0 & -\frac{1}{2}
\end{pmatrix}
\end{equation*}
\section*{Appendix C}
\noindent
Isospectral classes of the set $\fE_{\cJ}$:\\[4mm]
\textbf{I.} The class $\fE_{1}$ containing $32$ density matrices
$\ro$ with the eigenvalues (recall that $\mu_{k}^{(\alpha_{k})}$ denotes  eigenvalue $\mu_{k}$ with multiplicity $\alpha_{k}$)
\begin{equation*}
\left(\frac{3-10\,t}{27}\right)^{(1)},\quad \left(\frac{3-4\,t}{27}\right)^{(3)},\quad
\left(\frac{3+2\,t}{27}\right)^{(3)},\quad \left(\frac{3+8\,t}{27}\right)^{(2)}
\end{equation*}
and $t\in\,[-3/8,\, 3/10]$. Now $\fE_{1}=\bigcup\limits_{p=1}^{8}\fE_{1,p}$ and the classes $\fE_{1,p}$
are given by the following equivalence classes in $\cJ_{8}$:\\[4mm]
\textbf{1.}  $[I_{1,1}],\quad I_{1,1}=\mr{diag}\, (1,1,1,1,1,1,1,-1)$\\[2mm]
\textbf{2.} $[I_{1,2}],\quad I_{1,2}=\mr{diag}\,(1,1,1,1,1,-1,-1,-1)$\\[2mm]
\textbf{3.} $[I_{1,3}],\quad I_{1,3}=\mr{diag}\,(1,1,-1,1,1,1,-1,-1)$\\[2mm]
\textbf{4.} $[I_{1,4}],\quad I_{1,4}=\mr{diag}\, (1,1,-1,1,1,-1,1,-1)$\\[2mm]
\textbf{5.} $[I_{1,5}],\quad I_{1,5}=\mr{diag}\, (1,1,-1,1,-1,1,1,-1)$\\[2mm]
\textbf{6.} $[I_{1,6}],\quad I_{1,6}=\mr{diag}\, (1,1,-1,-1,1,1,1,-1)$\\[2mm]
\textbf{7.} $[I_{1,7}],\quad I_{1,7}=\mr{diag}\,(1,-1,-1,1,1,1,1,-1)$\\[2mm]
\textbf{8.} $[I_{1,8}],\quad I_{1,8}=\mr{diag}\,(-1,1,-1,1,1,1,1,-1)$\\[4mm]
\textbf{II.} The class $\fE_{2}$ containing $16$ density matrices
$\ro$ with the eigenvalues
\begin{equation*}
\left(\frac{3-10\,t}{27}\right)^{(1)},\quad  \left(\frac{3-4\,t}{27}\right)^{(1)},
\left(\quad \frac{3+2\,t}{27}\right)^{(4)},\quad
\left(\frac{2+8\,t}{27}\right)^{(1)},\quad
\left(\frac{3-(1+3\sqrt{5})\,t}{27}\right)^{(1)},\quad
\left(\frac{3-(1-3\sqrt{5})\,t}{27}\right)^{(1)}
\end{equation*}
and $t\in [-3/8,\, 2/10]$. In this case
$\fE_{2}=\bigcup\limits_{p=1}^{4}\fE_{2,p}$ and the classes
$\fE_{2,p}$ are given by the following equivalence classes in
$\cJ_{8}$:\\[4mm]
\textbf{1.} $[I_{2,1}],\quad I_{2,1}=\mr{diag}\,
(1,1,1,1,1,1,-1,-1)$\\[2mm]
\textbf{2.} $[I_{2,2}],\quad
I_{2,2}=\mr{diag}\,(1,1,1,1,1,-1,1,-1)$\\[2mm]
\textbf{3.} $[I_{2,3}],\quad
I_{2,3}=\mr{diag}\,(1,1,1,1,-1,1,1,-1)$\\[2mm]
\textbf{4.} $[I_{2,4}],\quad
I_{2,4}=\mr{diag}\,(1,1,1,-1,1,1,1,-1)$\\[4mm]
\textbf{III.} The class $\fE_{3}$ containing $16$ density matrices
$\ro$ with the eigenvalues
\begin{equation*}
\left(\frac{3-8\,t}{27}\right)^{(3)},\quad \text{and}\quad \left(\frac{3+4\,t}{27}\right)^{(6)}
\end{equation*}
and $t\in [-3/4,\, 3/8]$. Now
$\fE_{3}=\bigcup\limits_{p=1}^{4}\fE_{3,p}$ and the classes
$\fE_{3,p}$ correspond to:\\[2mm]
\textbf{1.} $[\I_{8}]$\\[2mm]
\textbf{2.} $I_{3,2}=\mr{diag}\,(1,1,1,1,1,-1,-1,1)$\\[2mm]
\textbf{3.} $I_{3,3}=\mr{diag}\,(1,-1,-1,1,1,1,1,1)$\\[2mm]
\textbf{4.} $I_{3,4}=\mr{diag}\,(-1,1,-1,1,1,1,1,1)$\\[4mm]
\textbf{IV.} The class $\fE_{4}$ containing $28$ density matrices
$\ro$ with the eigenvalues
\begin{equation*}
\left(\frac{3-8\,t}{27}\right)^{(1)},\quad \left(\frac{3-2\,t}{27}\right)^{(4)},\quad
\left(\frac{3+4\,t}{27}\right)^{(2)},\quad
\left(\frac{3-(6\sqrt{2}-4)\,t}{27}\right)^{(1)},\quad
\left(\frac{3+(6\sqrt{2}+4)\,t}{27}\right)^{(1)}
\end{equation*}
and $t\in [-3/(6\sqrt{2}+4),\, 3/8]$. In this case
$\fE_{4}=\bigcup\limits_{p=1}^{7}\fE_{4,p}$ and the classes
$\fE_{4,p}$ are given by:\\[2mm]
\textbf{1.} $[I_{4,1}],\quad
I_{4,1}=\mr{diag}\,(1,1,1,1,-1,1,-1,1)$\\[2mm]
\textbf{2.} $[I_{4,2}],\quad
I_{4,2}=\mr{diag}\,(1,1,1,1,-1,-1,1,1)$\\[2mm]
\textbf{3.} $[I_{4,3}],\quad
I_{4,3}=\mr{diag}\,(1,-1,1,1,1,1,-1,1)$\\[2mm]
\textbf{4.} $[I_{4,4}],\quad I_{4,4}=\mr{diag}\,(1,-1,1,1,1,-1,1,1)$\\[2mm]
\textbf{5.} $[I_{4,5}],\quad
I_{4,5}=\mr{diag}\,(1,-1,1,1,-1,1,1,1)$\\[2mm]
\textbf{6.} $[I_{4,6}],\quad
I_{4,6}=\mr{diag}\,(1,-1,1,-1,1,1,1,1)$\\[2mm]
\textbf{7.} $[I_{4,7}],\quad
I_{4,7}=\mr{diag}\,(1,-1,-1,1,-1,1,-1,1)$\\[4mm]
\textbf{V.} The class $\fE_{5}$ containing $12$ density matrices
$\ro$ with the eigenvalues
\begin{equation*}
\left(\frac{3-10\,t}{27}\right)^{(2)},\quad \left(\frac{3+2\,t}{27}\right)^{(6)},\quad
\left(\frac{3+8\,t}{27}\right)^{(1)}
\end{equation*}
and $t\in [-3/8,\,3/10]$. In this case
$\fE_{5}=\bigcup\limits_{p=1}^{3}$ and the classes $\fE_{5,p}$ are
given by:\\[2mm]
\textbf{1.} $[I_{5,1}],\quad
I_{5,1}=\mr{diag}\,(1,1,1,1,-1,1,-1,-1)$\\[2mm]
\textbf{2.} $[I_{5,2}],\quad
I_{5,2}=\mr{diag}\,(1,1,1,1,-1,-1,1,-1)$\\[2mm]
\textbf{3.} $[I_{5,3}],\quad
I_{5,3}=\mr{diag}\,(1,-1,-1,1,-1,1,-1,-1)$\\[4mm]
\textbf{VI.} The class $\fE_{6}$ containing $16$ density matrices $\ro$ with the eigenvalues
\begin{equation*}
\left(\frac{3-4\,t}{27}\right)^{(3)},\quad \left(\frac{2+3\, t}{27}\right)^{(2)},\quad \left(\frac{3
+8\, t}{27}\right)^{(1)},\quad \left(\mu_{4} (t)\right)^{(1)},\quad
\left(\mu_{5}(t)\right)^{(1)},\quad \left(\mu_{6}(t)\right)^{(1)}
\end{equation*}
where the eigenvalues $\mu_{4}(t),\, \mu_{5}(t),\,\mu_{6}(t)$ are given only numerically, and $t\in [t_{1},\, t_{2}]$
with $t_{i}$ satisfying $\mu_{4}(t_{i})=0$. One can check that $t_{1}\approx -0.3163$ and $t_{2}\approx 0.3404$. In this case
$\fE_{6}=\bigcup\limits_{p=1}^{4}\fE_{6,p}$ and the classes $\fE_{6,p}$ are given by;\\[2mm]
\textbf{1.} $[I_{6,1}],\quad I_{6,1}=\mr{diag}\,(1,-1,1,1,1,1,-1,-1)$\\[2mm]
\textbf{2.} $[I_{6,2}],\quad I_{6,2}=\mr{diag}\,(1,-1,1,1,1,-1,1,-1)$\\[2mm]
\textbf{3.} $[I_{6,3}],\quad I_{6,3}=\mr{diag}\,(1,-1,1,1,-1,1,1,-1)$\\[2mm]
\textbf{4.} $[I_{6,4}],\quad I_{6,4}=\mr{diag}\,(1,-1,1,-1,1,1,1,-1)$\\[4mm]
\textbf{VII.} The class $\fE_{7}$ containing $4$ locally equivalent density matrices $\ro$ with the eigenvalues
\begin{equation*}
\left(\frac{3-2\, t}{27}\right)^{(8)} \quad \text{and}\left(\frac{3+16\, t}{27}\right)^{(1)}
\end{equation*}
and $t\in [-3/16,\, 3/2]$. In this case there is only one equivalence class $[I_{7,1}]$ defined by
$I_{7,1}=I_{0}=\mr{diag}\,(1,-1,1,1,-1,1,-1,1)$.\\[4mm]
\textbf{VIII.} The class $\fE_{8}$ containing $4$ locally equivalent density matrices $\ro$ with the eigenvalues
\begin{equation*}
\left(\frac{3-4\,t}{27}\right)^{(3)},\quad \left(\frac{3+2\,t}{27}\right)^{(4)},\quad
\left(\frac{3-(6\sqrt{3}-2)\,t}{27}\right)^{(1)},\quad
\left(\frac{3+(6\sqrt{3}+2)\,t}{27}\right)^{(1)}
\end{equation*}
and $t\in [-3/(6\sqrt{3}+2),\, 3/(6\sqrt{3}-2)]$. In this case there
is only one equivalence class $[I_{8,1}]$ defined by
$I_{8,1}=\mr{diag}\,(1,-1,1,1,-1,1,-1,-1)$.\\[4mm]
Now we discuss entanglement properties of the states from the set
$\fE_{\cJ}$.  To detect entangled states of two qutrits we apply Peres - Horodecki criterion of separability:
all separable states are positive under partial transposition (PPT states) and when the partial transposition is non - positive
(NPPT states) such states are entangled. To measure entanglement of the state we
use its negativity $N(\ro)$. Negativity
 is an entanglement monotone, but it cannot detect entangled
states that are positive under partial transpose (bound entangled
PPT states). To detect some of the bound entangled PPT states, we
can use the realignment criterion of separability \cite{CW}. This
criterion states that for any separable state $\ro$, the matrix
$\ro^{\mr{R}}$ with elements
\begin{equation*}
\bra{m}\otimes\bra{\mu}\,\ro^{\mr{R}}\,\ket{n}\otimes\ket{\nu}=\bra{m}\otimes
\bra{n}\,\ro\,\ket{\mu}\otimes\ket{\nu}
\end{equation*}has a trace norm not greater then $1$. So if
realignment negativity defined by
\begin{equation*}
N_{\mr{R}}(\ro)=\max\,\left(0,\,
\frac{||\ro^{\mr{R}}||_{1}-1}{2}\right)
\end{equation*}
is non zero, the state $\ro$ is entangled and in the case when $N(\ro)=0$, it is bound entangled.
\par
Let us apply the measures $N$ and $N_{\mr{R}}$ to the states from
$\fE_{\cJ}$. It follows that all states in the classes $\fE_{4}$ and
$\fE_{5}$  have zero negativities $N$ and $N_{\mr{R}}$,
and it suggests that $\fE_{4}$ and $\fE_{5}$ contain only separable
states. On the other hand, the remaining states have non - zero
negativity $N$, for some values of the parameter $t$. In particular:\\[4mm]
\textbf{a.} $\ro\in \fE_{1,1}$ are NPPT for $-3/8\leq\, t \,<-3/a$ and PPT for $-3/a \leq\, t\,\leq 3/10,\;
a=2+6\sqrt{3}$,\\[2mm]
\textbf{b.}   $\ro\in \fE_{1,k}, k=2,\ldots,8$ are PPT for $-3/8\,\leq t\,\leq 3/b$ and NPPT for
$3/b<\,t\,\leq 3/10,\; b=4+6\sqrt{2}$,\\[2mm]
\textbf{c.}  $\ro\in \fE_{2}$ are NPPT for $-3/8\leq \,t <\,t_{0}$ and PPT for $t_{0}\leq\, t \leq 2/10,\;
t_{0}\approx -0.316$,\\[2mm]
\textbf{d.} $\ro\in \fE_{3,1}$ are NPPT for $-3/4 \leq\,t\,<-3/16$ and PPT for $-3/16\leq\,t\,\leq 3/8$,\\[2mm]
\textbf{e.} $\ro\in \fE_{3,k},\, k=2,3,4$ are NPPT for
$-3/4\leq\,t\,< -3/10$ and PPT for $-3/10\leq\,t\,\leq 3/8$,\\[2mm]
\textbf{f.} $\ro\in \fE_{6}$ are PPT for $t_{1}\leq\,t\leq \,
3/10,\, t_{1}\approx -0.3163$ and NPPT for $3/10<\,t\,\leq t_{2},\,t_{2}\approx 0.3404$\\[2mm]
\textbf{g.} $\ro\in \fE_{8}$ are PPT for $-3/(6\sqrt{3}+2)\leq\,t\leq\,3/10$ and NPPT
for $3/10<\,t\,\leq 3/(6\sqrt{3}-2)$.\\[4mm]
The class $\fE_{7}$ is distinguished for many reasons. It is the
only class in $\fE_{\cJ}$ which contains pure states and those
states are maximally entangled. The states $\ro\in \fE_{7}$ are PPT
when $-3/16\leq t\leq 3/8$ and NPPT, when $3/8< t \leq 3/2$. All
NPPT states in this class are free entangled i.e. they are
distillable \cite{BB}. This follows from the fact that all NPPT states from the class $\fE_{7}$
violate  reduction
criterion of separability \cite{CAG}
\begin{equation*}
\ptr{B}\ro\otimes \I-\ro\geq 0\quad\text{and}\quad \I\otimes
\ptr{A}-\ro\geq 0
\end{equation*}
so are distillable \cite{HH}.

\end{document}